\documentclass{aa}
\usepackage{graphicx}
\usepackage{natbib,twoopt}
\usepackage{lscape}
\usepackage{rotating}
\usepackage[breaklinks=true]{hyperref}
\usepackage{mathtools}
\bibpunct{(}{)}{;}{a}{}{,}
\usepackage[varg]{txfonts}

\begin{document}

\title{The nature of the low-frequency emission of M51\thanks{The total intensity FITS file is only available in electronic form
at the CDS via anonymous ftp to cdsarc.u-strasbg.fr (130.79.128.5) or via http://cdsweb.u-strasbg.fr/cgi-bin/qcat?J/A+A/}}
\subtitle{First observations of a nearby galaxy with LOFAR}

\author{D.\,D.~Mulcahy\inst{1}\fnmsep\thanks{Now at the School of Physics and Astronomy, University of Southampton, Highfield, SO17 1SJ, Southampton, UK; d.d.mulcahy@soton.ac.uk}
\and A.~Horneffer\inst{1}
\and R.~Beck\inst{1}
\and G.~Heald\inst{2,3}
\and A.~Fletcher\inst{4}
\and A.~Scaife\inst{5}
\and B.~Adebahr\inst{1}
\and J.\,M.~Anderson\inst{1}\fnmsep\thanks{Now at the Leibniz-Institut f\"{u}r Astrophysik Potsdam, An der Sternwarte 16, 14482 Potsdam, Germany}
\and A.~Bonafede\inst{6}
\and M.~Br\"{u}ggen\inst{6}
\and G.~Brunetti\inst{7}
\and K.\,T.~Chy\.zy\inst{8}
\and J.~Conway\inst{9}
\and R-J.~Dettmar\inst{10}
\and T.~En{\ss}lin\inst{11}
\and M.~Haverkorn\inst{12,13}
\and C.~Horellou \inst{9}
\and M.~Iacobelli\inst{2,13}
\and F.\,P.~Israel\inst{13}
\and H.~Junklewitz\inst{14}
\and W.~Jurusik\inst{8}
\and J.~K\"{o}hler\inst{1}
\and M.~Kuniyoshi\inst{1}
\and E.~Orr\'{u}\inst{2}
\and R.~Paladino\inst{15,7}
\and R.~Pizzo\inst{2}
\and W.~Reich\inst{1}
\and H.\,J.\,A.~R\"{o}ttgering\inst{13}}

\institute{Max-Planck-Institut f\"{u}r Radioastronomie, Auf dem H\"{u}gel 69, 53121 Bonn, Germany
\and ASTRON, Postbus 2, 7990 AA, Dwingeloo, The Netherlands
\and Kapteyn Astronomical Institute, Postbus 800, 9700 AV Groningen, The Netherlands
\and School of Mathematics and Statistics, Newcastle University, Newcastle-upon-Tyne NE1 7RU, UK
\and School of Physics and Astronomy, University of Southampton, Highfield, SO17 1SJ, Southampton, UK
\and Universit\"{a}t Hamburg Sternwarte, Gojenbergsweg 112, 21029 Hamburg, Germany
\and INAF-IRA Bologna, Via Gobetti 101, 40129 Bologna, Italy
\and Astronomical Observatory, Jagiellonian University, ul. Orla 171, 30-244 Krak\'ow, Poland
\and Dept. of Earth \& Space Sciences, Chalmers University of Technology, Onsala Space Observatory, 439 92 Onsala, Sweden
\and Ruhr-Universit\"{a}t Bochum, Astronomisches Institut, 44780 Bochum, Germany
\and Max-Planck-Institut f\"{u}r Astrophysik, Karl-Schwarzschild-Str. 1, 85748 Garching, Germany
\and Department of Astrophysics/IMAPP, Radboud University Nijmegen, P.O. Box 9010, 6500 GL Nijmegen, The Netherlands
\and Leiden Observatory, Leiden University, PO Box 9513, 2300 RA Leiden, the Netherlands
\and Argelander-Institut f\"{u}r Astronomie, Radio Astronomy Department, Auf dem H\"{u}gel 71, 53121 Bonn, Germany
\and Department of Physics and Astronomy, University of Bologna, V.le Berti Pichat 6/2, 40127 Bologna, Italy
}

\date{Received 12 May 2014 /
Accepted 1 July 2014}

\abstract{Low-frequency radio continuum observations (< 300\,MHz) can provide valuable information on the propagation of low-energy cosmic ray electrons. Nearby spiral galaxies have hardly been studied in this frequency range because of the technical challenges of low-frequency radio interferometry. This is now changing with the start of operations of LOFAR.}
{We aim to study the propagation of low-energy cosmic ray electrons in the interarm regions and the extended disk of
the nearly face-on spiral galaxy Messier 51. We also search for polarisation in M51 and other extragalactic sources in the field.}
{The grand-design spiral galaxy M51 was observed with the LOFAR High Frequency Antennas (HBA) and imaged
in total intensity and polarisation. This observation covered the frequencies
between 115\,MHz and 175\,MHz with 244 subbands of 8 channels each, resulting in 1952 channels. This allowed us
to use RM synthesis to search for polarisation.}
{We produced an image of total emission of M51 at the mean frequency of 151\,MHz with 20\arcsec\ resolution and 0.3\,mJy
rms noise, which is the most sensitive image of a galaxy at frequencies below 300\,MHz so far. The integrated spectrum of total
radio emission is described well by a power law, while flat spectral indices in the central region indicate thermal absorption.
We observe that the disk extends out to 16\,kpc and see a break in the radial profile near the optical radius of the disk.
The radial scale lengths in the inner and outer disks are greater at 151\,MHz, and the break is smoother at 151\,MHz
than those observed at 1.4\,GHz. The arm--interarm contrast is lower at 151\,MHz than at 1400\,MHz, indicating
propagation of cosmic ray electrons (CRE) from spiral arms into interarm regions.
The correlations between the images of radio emission at 151\,MHz and 1400\,MHz and the far-infrared emission at
$70\,\mu$m reveal breaks on scales of 1.4 and 0.7\,kpc, respectively.
The total (equipartition) magnetic field strength decreases from about $28\,\mu$G in the central region to about
$10\,\mu$G at 10\,kpc radius.
No significant polarisation was detected from M51, owing to severe Faraday depolarisation. Six extragalactic sources
are detected in polarisation in the M51 field of $4.1\degr \times 4.1\degr$ size. Two sources show complex
structures in Faraday space.}
{Our main results, the scale lengths of the inner and outer disks at 151\,MHz and 1.4\,GHz, arm--interarm contrast, and
the break scales of the radio--far-infrared correlations, can be explained consistently by CRE diffusion, leading
to a longer propagation length of CRE of lower energy. The distribution of CRE sources drops sharply at about 10\,kpc
radius, where the star formation rate also decreases sharply. We find evidence that thermal absorption is primarily caused by {\sc H\,ii} regions.
The non-detection of polarisation from M51 at 151\,MHz is consistent with the estimates of Faraday depolarisation.
Future searches for polarised emission in this frequency range should concentrate on regions with low star formation rates.}


\maketitle

\section{Introduction}

Messier 51, or NGC5194, aptly named the "Whirlpool Galaxy", is one of the most widely studied galaxies because it is one of the most striking nearby examples of a classical grand-design spiral galaxy.
Spiral galaxies exhibit a wide range of morphologies, from flocculent to grand-design, that can be caused by small-scale gravitational instabilities or caused by larger scale perturbations caused by global density waves, tidal interactions, or bars.
In the case of M51, many numerical simulations have been performed to study the consequences of the gravitational interaction between the main galaxy, NGC5194, and its nearby companion NGC5195 \citep{2000MNRAS.319..377S,2003Ap&SS.284..495T}.
For example, \citet{2010MNRAS.403..625D} used hydrodynamical models to simulate the tidally induced spiral structure. In these studies the authors were able to replicate several features seen in observations, such as the {\sc H\,i} tidal tail \citep{1990AJ....100..387R} and kinks and bifurcations in the spiral arms. The physical parameters of M51 used in this paper can be seen in Table~\ref{physicalpara}.

M51 is the first external galaxy where polarised radio emission was detected with the Westerbork Synthesis Radio Telescope, WSRT
\citep{1972A&A....17..468M}. polarised emission is observed throughout most of the disk of M51, as observed at
10.7\,GHz with the Effelsberg 100-m telescope by \citet{1992A&A...263...30N}, and the pattern of magnetic field
lines follows that of the optical spiral arms. \citet{1992A&A...265..417H} found from VLA observations at frequencies
of 1.47 and 1.67\,GHz that the distribution of polarised emission is strongly affected by Faraday depolarisation and
that the polarised emission is actually emerging from an upper layer of the disk, because the galaxy is not
transparent to polarised emission at these frequencies.
\citet{2009A&A...503..409H} again observed polarisation from M51 at 1.30--1.76\,GHz throughout the whole disk; however, it was seen that there were large variations with a 5\% polarisation fraction detected in the optical
disk and 25--30\%  beyond the outer arms.
\citet{2009A&A...503..409H} also performed RM Synthesis on these data and detected a main Faraday component at +13\,rad\,m$^{-2}$, with the Faraday rotation dominated by Milky Way foreground.
\citet{2010A&A...514A..42B} made additional use of this technique and studied two weaker secondary components near $-$180 and +200\,rad\,m$^{-2}$, coming from M51 itself.

\begin{table}[h!]
\caption{Physical parameters of M51=NGC5194}
\centering
\begin{tabular}{l l}
\hline\hline
Morphology & SAbc\\
Position of the nucleus & $\alpha(2000)=13^\mathrm{h} 29^\mathrm{m} 52^\mathrm{s}.709$\\
 & $\delta(2000)=+47^\circ 11^\prime 42.59^{\prime \prime}$ \\
Position angle of major axis & $-10^\circ$ ($0^\circ$ is North)\\
Inclination & $-20^\circ$ ($0^\circ$ is face on) \tablefootmark{ a} \\
Distance & 7.6 Mpc \tablefootmark{ b} \\
Optical radius ($\mathit{R}_{25}$) & 3.9$^{\prime}$ (8.6 kpc) \tablefootmark{ c} \\
\hline
\end{tabular}
\tablefoottext{a}{\citet{1974ApJS...27..437T}}
\tablefoottext{b}{\citet{2002ApJ...577...31C}}
\tablefoottext{c}{\citet{2010AJ....140.1194B}}
\label{physicalpara}
\end{table}

High-resolution VLA observations at 4.86\,GHz, combined with Effelsberg data at similar frequencies
\citep{2011MNRAS.412.2396F}, have shown that the locations of the inner radio spiral arms coincide with CO emission,
while the polarised emission peaks at the inner edge of the material arms \citep{2006A&A...458..441P}. The polarised emission mostly emerges from anisotropic turbulent fields generated by compressing and shearing gas flows.
\citet{2013ApJ...779...42S} found evidence in M51 for a physical link between the molecular gas as observed in
the CO line and the total magnetic field seen in the total radio continuum emission, both at 2\arcsec\ resolution.

From the analysis of VLA plus Effelsberg polarisation observations at 4.86 and 8.46\,GHz, \citet{2011MNRAS.412.2396F}
were able to identify two underlying patterns of regular magnetic fields. A regular field with a combination of
$m=0+2$ azimuthal modes was found in the disk, while in the halo a regular field with a bisymmetric
$m=1$ azimuthal mode predominates. The origin of this regular field is probably due to two large-scale dynamos
\citep{1996ARA&A..34..155B} operating in the disk and the halo.

Considering how extensively M51 has being studied both observationally and theoretically, very few observations
have being performed on M51 at frequencies below 500\,MHz, primary due to the difficulties of observing at low radio
frequencies.
Low-frequency radio synchrotron emission is very important as it is produced by aged and low-energy electrons
that are less affected by energy losses and therefore can propagate further away from their site of origin
(assuming that the diffusion coefficient does not vary with particle energy).
Depending on the magnetic field strength in the outer disk, one would expect to see a large radio synchrotron disk around M51.
The extent and profile of the radio disk at low frequencies contains information about the propagation of
cosmic ray electrons and the strength of the magnetic field beyond the disk of strong star formation. This will be
investigated in this paper.
\citet{1977A&A....54..703S} observed M51 with the WSRT at 610\,MHz as well as 1.4\,GHz, but with a
resolution of approximately $1^{\prime}$ for the 610\,MHz observation, only few details of the inner galaxy could
be observed. An extended component could be seen but this extended no further than the 1.4\,GHz observation.
Using both observations \citet{1977A&A....54..703S} computed a spectral index image which was compared with analytical
solutions of the cosmic ray diffusion equation.
Two sets of parameters produced models which could fit the observations, namely: (a) a constant magnetic
field and diffusion coefficient with leakage of electrons out of the galactic magnetic field perpendicular to the disk,
 or (b) no leakage but a magnetic field that is proportional to $1/r^{1/2}$ and a diffusion coefficient
that is proportional to $1/r$, where $r$ is the galactocentric distance.
However, the step-like source function used in the analytical solutions is now outdated and unrealistic.

M51 was also observed at low frequencies as part of a survey of 133 spiral galaxies performed by \citet{1990ApJ...352...30I} (hereafter I \& M) using the Clarke Lake Telescope at 57.5\,MHz with a total flux density of $11\pm1.5$\,Jy. However these results may suffer from uncertainties. Overall,
the measured total flux densities in nearly all galaxies of this survey were lower than expected through extrapolating
from higher frequencies. In addition,
it was found that more highly inclined galaxies had lower fluxes than moderate-inclined galaxies. I\&M explained this flattening of the radio spectrum through free-free absorption,
especially because of its dependence on inclination. Such an effect may indicate the
existence of a clumpy medium of well-mixed, non-thermally emitting and thermally absorbing gas with a small filling factor and an electron temperature of 500 to 1000\,K.
\cite{1991A&A...251..442H} reanalysed the same data and confirmed the spectral index flattening, but found that the magnitude
of the flattening does not depend on the inclination of the galaxy.
This is one of the main issues that needs to be addressed with low-frequency observations, especially with the upcoming Multifrequency Snapshot Sky Survey (MSSS) \citep{NewEntry6}.

Little is known about the conditions in the ISM of the outer disk
($r \textgreater R_{25}$). \citet{2006ApJ...651L.101T}, using IR, UV, H$\alpha$ and {\sc H\,i} measurements, found that
star formation in the outer disk of M51 is about an order of magnitude less efficient than at smaller radii.  \citet{2004Natur.432..369B} observed carbon monoxide (CO) in the outer disk of NGC4414 in regions of high {\sc H\,i}
column density regions. In these regions, \citet{2010AJ....140.1194B} found that despite widespread
star formation, the outer disk star formation is extremely inefficient, that is, the gas depletion time is very low and
increases with radius. A low rate of star formation means that the level of turbulence is probably too low
to maintain a strong magnetic field \citep{1996ARA&A..34..155B}. On the other hand, turbulence could also be induced by the magneto-rotational
instability \citep{1999ApJ...511..660S} in the outer regions of the galaxy, thereby increasing the turbulent energy
density, magnetic field strength, and hence the scale length of radio synchrotron emission \citep{2007A&A...470..539B}.

We shall outline the results from the first LOFAR \citep{2013A&A...556A...2V} observations of a nearby
galaxy, namely M51. In Section~\ref{section2} we will
describe the observational setup along with the data reduction process. In Section~\ref{section3} we will present the LOFAR HBA 151\,MHz image of M51 and investigate the radial profile of LOFAR emission while comparing to higher frequencies. In Sections~\ref{section4} and \ref{section5} we will present and discuss the spectral index and total magnetic field strength images derived from the LOFAR 151\,MHz and VLA 1.4\,GHz image from \citet{2011MNRAS.412.2396F}.
In Section~\ref{section6}, the cosmic ray propagation in M51 will be discussed, and the wavelet cross correlation
between radio and far-infrared emission will be shown in Section~\ref{section7}.
In Section~\ref{section8} we shall also present results from performing rotation measure synthesis on this region and the analysis of detected background sources in the field. Finally, conclusions and prospects for future work will be presented in Sections~\ref{section9} and \ref{section10}.

\section{Observations and preprocessing data reduction \label{section2}}

The M51 field was observed for 8 hours with a configuration of 61 LOFAR High Band Antenna (HBA) fields --
48 core station fields and 13 remote station fields.
No international stations were used in this observation because sub-arcsec resolution was not needed and
adding them would have required the removal of some core stations.
The observation was done in dual beam mode, with one beam targeting M51
($\alpha=13^\mathrm{h} 29^\mathrm{m} 52^\mathrm{s}7$, $\delta=+47^\circ 11^\prime 43^{\prime \prime}$) and one beam targeting 3C295  ($\alpha=14^\mathrm{h} 11^\mathrm{m} 20^\mathrm{s}5$, $\delta=+52^\circ 12^\prime 10^{\prime \prime}$) simultaneously.
This could be done because 3C295 is less than 10 degrees from M51 and thus within the HBA analog tile beam which was centred on M51 itself.
The second beam was used for flux and initial phase calibration.
The so-called 8-bit mode\footnote{The data being sent by the LOFAR stations is encoded with 8-bit integers.}
of LOFAR was used, giving an instantaneous bandwidth of about 95\,MHz.
This total bandwidth is split into 488 subbands, each with a bandwidth of 0.195\,MHz. In this
observation the subbands were divided evenly between the calibrator (244 subbands) and the target (244 subbands). As these 244 subbands cannot fully cover the HBA band, the bandwidth in each beam was split into 8 blocks of approximately 30 subbands (about 6\,MHz bandwidth) spread evenly over the HBA band.
The observation was done between 115\,MHz and 175\,MHz to avoid the band edge of the analog filters at the low
frequencies and the the strong RFI\footnote{Radio Frequency Interference} from DAB\footnote{Digital Audio Broadcast}
above 175\,MHz \citep{2013A&A...549A..11O}. Details of the observations are given in Table~\ref{observationpara}.

\begin{table}[h!]
\caption{Parameters of the M51 LOFAR observations}
\centering
\begin{tabular}{l l}
\hline\hline
Start date (UTC) & 22-Apr-2013/ 20:08:07.0 \\
End date (UTC) & 23-Apr-2013/ 04:04:04.7 \\
\hline
Frequency range & 115.9--175.8 MHz  \\
Total bandwidth on target & 47.7 MHz  \\
Total bandwidth on calibrator & 47.7 MHz \\
\hline
\label{observationpara}
\end{tabular}
\end{table}

The majority of the data reduction was performed with the LOFAR pipeline. A full description of this pipeline is out of the scope of this paper. Interested readers can refer to  \citet{2010arXiv1008.4693H} and \citet{2013A&A...556A...2V}.

RFI excision was done with the AOFlagger pipeline \citep{2012A&A...539A..95O}, which was used on the
raw data and all linear correlations to give  the best results for RFI detection and flagging. The visibility data were checked to see if any contamination from the so called A-Team (Cassiopeia A, Cygnus A, Virgo A, and Taurus A) entered though the sidelobes. Only minimal contamination was seen, so "demixing" \citep{2007ITSP...55.4497V} was not
applied, but affected visibilities were flagged.
The data were then compressed to 8 channels per subband in frequency and 14 seconds sampling time. These flagging and
compression operations were performed using the New Default Pre-Processing Pipeline (NDPPP), which is part of the
LOFAR software.

\subsection{Initial calibration}

The calibrator subbands of 3C295 were calibrated using a sky model from \citet{2012MNRAS.423L..30S} using the
Black Board Selfcal (BBS) software \citep{2009ASPC..407..384P}.
Note that the flux scale is that of \citet{1973AJ.....78.1030R} (hereafter RBC). This scale is used to avoid the suggested issues \citep{1990MNRAS.244..233R} with the secular decrease
in the flux density of Cassiopeia A at low frequencies existing in the \citet{1977A&A....61...99B} scale.

The calibration solutions found from calibrating the 3C295 data were directly transferred to the corresponding
target data at the same frequency. This included the phase solutions as it made the next phase only calibration easier.

The target subbands for each block of approximately 30 subbands were combined in frequency
for a better signal to noise ratio. These blocks were then phase calibrated using the preliminary LOFAR global sky model.
This global sky model consists of positions of sources from the NRAO VLA Sky Survey (NVSS) catalogue \citep{1998AJ....115.1693C} with
fluxes obtained from a power-law spectral fitting with data from the NVSS, the Westerbork Northern Sky Survey (WENSS) \citep{1997A&AS..124..259R} and the VLA Low-Frequency Sky Survey redux
(VLSSr) \citep{2007AJ....134.1245C} \citep{2014MNRAS.440..327L} surveys.

After this initial calibration, each block of subbands was checked manually for any RFI or bad solutions resulting
from the calibration, which were in turn flagged.

\subsection{Self calibration}

Once the initial direction-independent calibration had been completed, a new sky model was created from images
produced by two programs.
First, an image of the full field-of-view was created using AWimager
\citep{2013A&A...553A.105T} which is part of the LOFAR software. This program utilises the A-projection
algorithm \citep{2008A&A...487..419B}, when imaging wide fields to correct for direction-dependent effects resulting
from changing primary beams and ionospheric effects. It can therefore yield accurate fluxes for sources far away
from the phase centre.
Next, a second image was created using CASA\footnote{http://casa.nrao.edu/} and utilising its multiscale clean
\citep{2008ISTSP...2..793C} and w-projection \citep{2005ASPC..347...86C} tasks solely for creating a detailed model
for M51 itself. Multiscale clean is needed to accurately image the extended structures in M51 and was not
available in the AWimager at the time of processing.

As M51 is very small compared to the size of the primary beam and is located at the phase centre of our observation,
errors due to the missing primary beam correction are small.
Both images were made with an uv coverage of 12\,k$\lambda$, which  corresponds to 20--25\arcsec\ resolution with
uniform weighting.
The clean components for M51 from the CASA image and the rest of the field from the AWimager were combined into
the new sky model. That was then used for direction-independent self calibration of the phases.
This whole process was performed on each of the 8 blocks of 30 subbands. No sources were subtracted from the visibilities during the calibration.

\begin{figure}
\resizebox{\hsize}{!}{\includegraphics{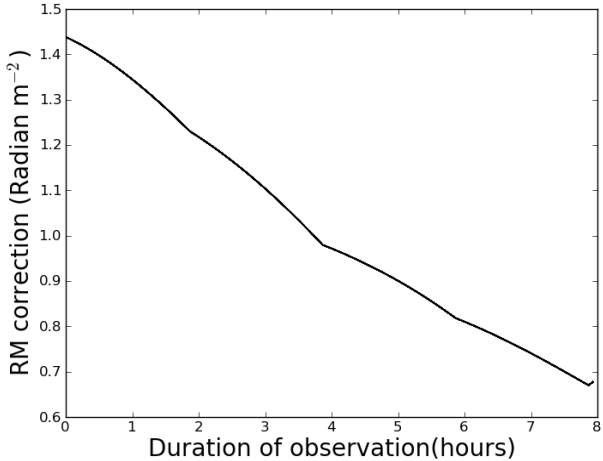}}
\caption{Calculated ionospheric RM correction in the direction of M51 for the duration of the observation.}
\label{rmcorr}
\end{figure}

At low frequencies, correction for ionospheric Faraday rotation is needed in order to avoid depolarisation caused
by the changing ionosphere.
Values of the RM correction were calculated using the method by \citet{2013A&A...552A..58S}.
Measurements of the vertical total electron content (VTEC) were obtained from the centre for Orbit Determination in
Europe (CODE) which have a time resolution of 2\,hours.
These were combined with the Earth's magnetic field as calculated using the eleventh generation of the International
Geomagnetic Reference Field (IGRF11; \citet{2010GeoJI.183.1216F-1}).
The calculated RM correction in the direction of M51 was applied to the data. The values for the entire observation
can be seen in Figure~\ref{rmcorr}.

\section{M51 total power image \label{section3}}

\begin{figure*}
\centering
\includegraphics[width=15cm]{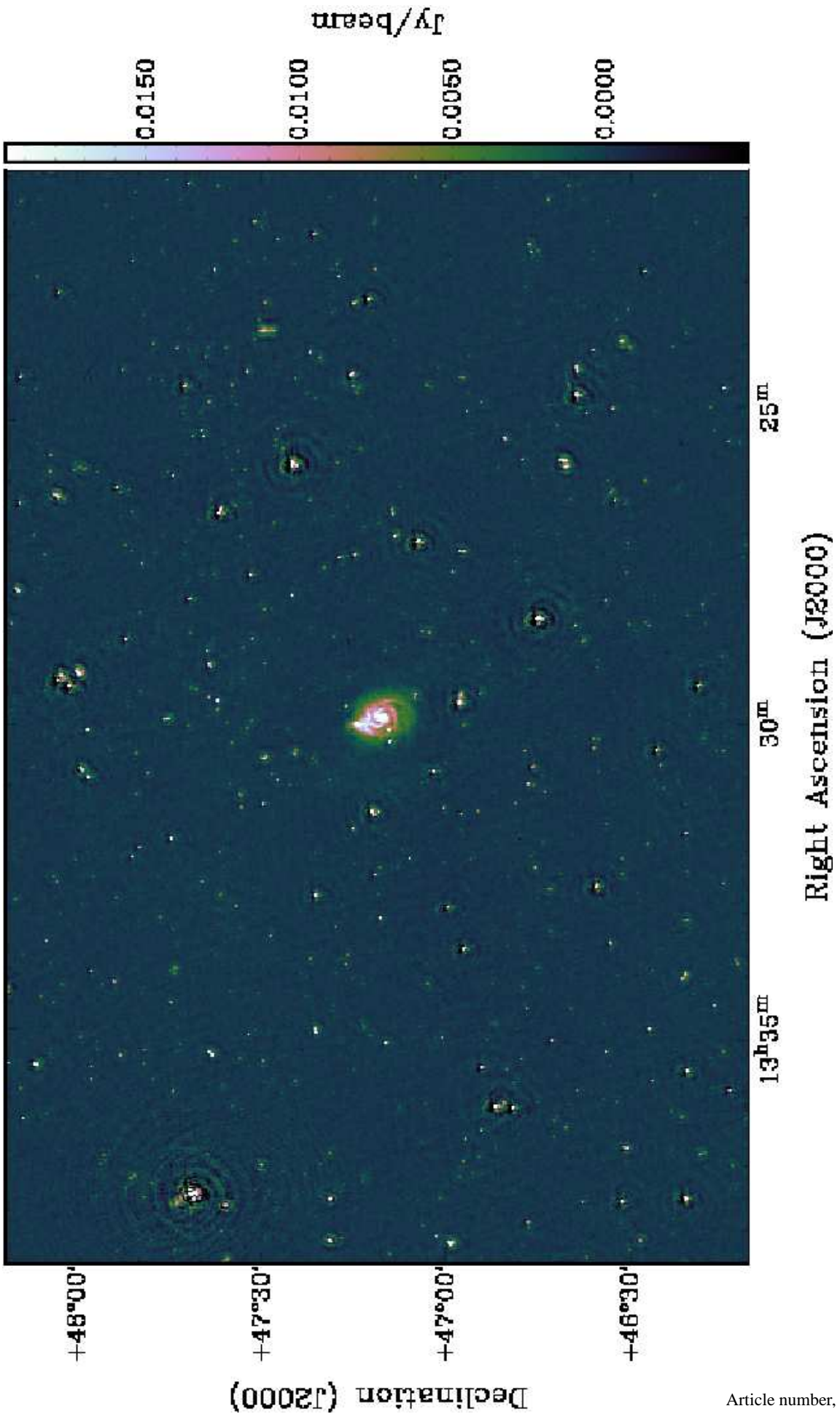} 
\caption{Region around M51 at a central frequency of 151\,MHz with a bandwidth of 47.7\,MHz. The resolution is 20\arcsec\, and a
robust weighting of $-0.5$ was used. The colour scale is in Jy/beam using a cubehelix colour scheme \citep{2011BASI...39..289G}.}
\label{widefieldimage}
\end{figure*}

Images were cleaned using the same imaging parameters and uv-distance (namely 16 k$\lambda$) for each of the eight subband
blocks with CASA in order to utilise multi-scale clean with w-projection. A mask was used in this cleaning which was created by hand.
Several point sources were compared to the point spread function (PSF) to investigate if the point sources were
broadened due to phase errors caused by the ionosphere, but
no significant broadening could be seen. Also, the astrometry of these point sources was checked with catalogue values,
but no major deviations could be seen.

\begin{figure*}
\centering
\includegraphics[width=16cm]{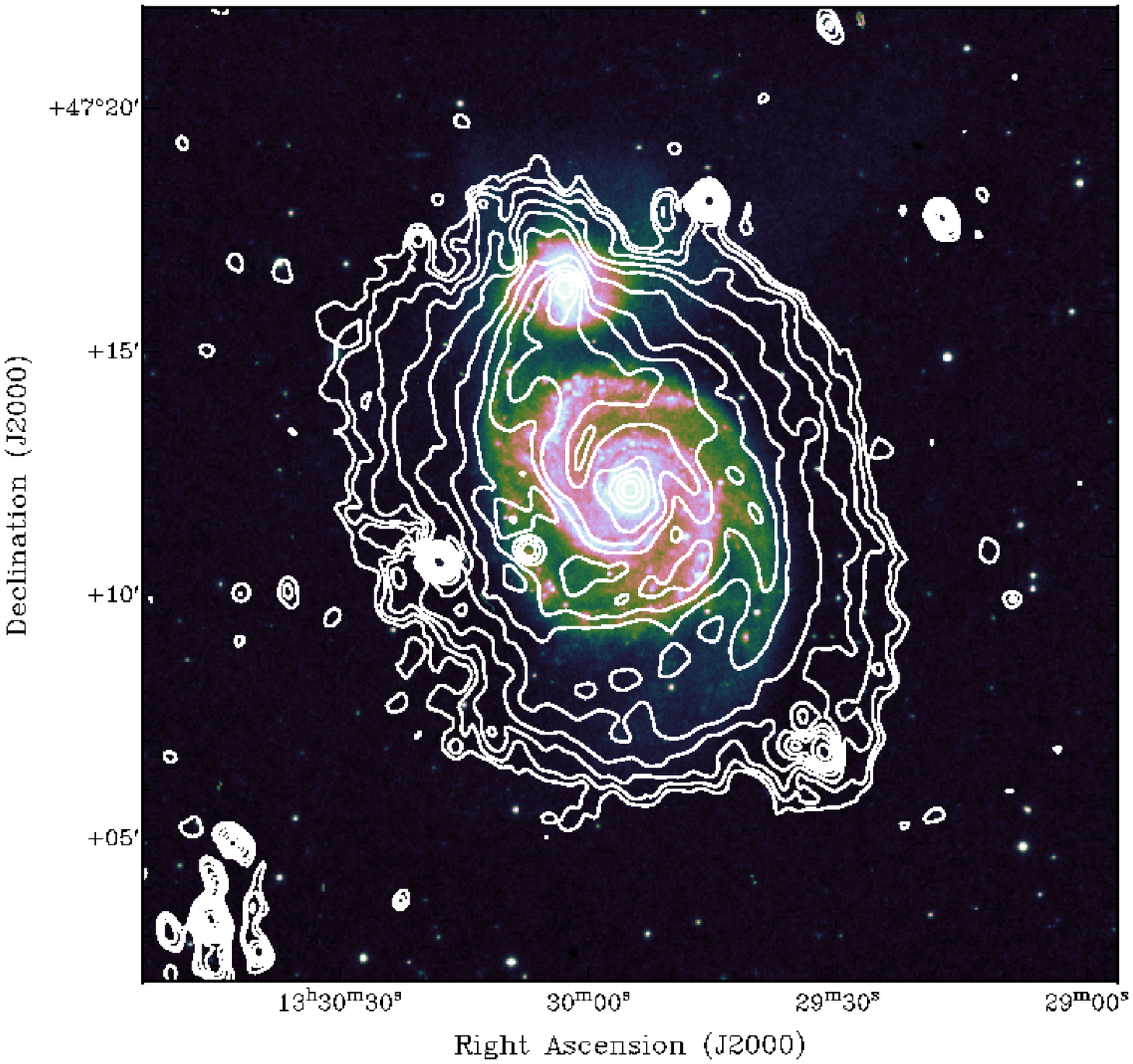}
\caption{M51 at a central frequency of 151\,MHz with a bandwidth of 47.7\,MHz overlayed onto an optical DSS image using a cubehelix colour scheme \citep{2011BASI...39..289G}.
The resolution is 20\arcsec\, and a robust weighting of $-0.5$ was used. Here the extended disk is seen clearly.
The contours start at 1 mJy/beam and increase by a factor 1.5.}
\label{contourplusopticalimage}
\end{figure*}

All eight frequency block images were averaged in the image plane using an
inverse variance weighting scheme. The central frequency is 151\,MHz. The final image of M51 at 20\arcsec\ resolution
with robust weighting of $-0.5$ is shown in Fig.~\ref{widefieldimage} and Fig.~\ref{contourplusopticalimage}
which shows the extended disk well.

The rms noise in quiet regions close to the edges of the field of view (FoV) for both uniform and robust weighting schemes is approximately 150\,$\mu$Jy/beam.
Closer to M51 the noise in the image with uniform weighting is between 200 and 300\,$\mu$Jy/beam while the noise in the robust image (Fig.~\ref{contourplusopticalimage}) is between 300 and 400\,$\mu$Jy/beam. The theoretical thermal noise value is expected to be about 30\,$\mu$Jy/beam or taking into account the weighting scheme, 60\,$\mu$Jy/beam.
This means we achieve approximately 2.5 -- 5 times the thermal noise.
This is the deepest image obtained so far for any galaxy in the low-frequency regime ($<$ 300\,MHz).
Phase errors are seen around the brighter sources in the field but these are localised.
No phase errors are obvious near M51.
Directional-dependent calibration would need to be performed in many directions to get rid of these phase errors completely. This will be done in a later work when studying M51 at much higher resolution.

The grand design spiral arms of M51 still can be clearly seen at 151\,MHz. The arm and interarm contrast is
highly visible in M51 at these low frequencies and comparable at first glance to higher frequencies
\citep{2011MNRAS.412.2396F}. This is in stark contrast to the galaxies observed at
333\,MHz with the GMRT where the spiral arms and inter arm regions were often indiscernible from each other \citep{2012MNRAS.419.1136B}.
According to \citet{2011MNRAS.412.2396F}, this is primarily caused by the low-energy cosmic ray electrons (CREs)
diffusing from star forming regions in the spiral arms without losing much energy compared to higher
energy electrons seen at higher frequencies. This may enable the lower energy electrons to diffuse further and thus fill up the interarm regions.

CRE diffusion from the spiral arms into the interarm region can be seen in the 151 MHz\,image when compared to
images at higher frequencies. Especially in the northern region, the interarm region is becoming
squeezed as the CRE diffuse into the interarm region. However, it is not as severe compared to the galaxies
observed by \citet{2012MNRAS.419.1136B}.

The region between the companion and the northern arm of NGC5194 where little IR emission exists \citep{2011AJ....141...41D-1} is very distinct at 151\,MHz. It is much brighter compared to the 4.86 and 8.46\,GHz images by \citet{2011AJ....141...41D-1} and \citet{2011MNRAS.412.2396F}, indicative
of synchrotron emission that is enhanced by the ongoing interaction. This is seen in Section~\ref{section4}
where the spectral index is larger than $-0.8$.

Supernova SN 2011dh, discovered on 01 June 2011 by the Palomar Transient Factory project (PTF)
\citep{2011ApJ...742L..18A}, \citep{2011ATel.3398....1S}, was not detected in our observations.
This gives an upper limit of the SN flux density at 151\,MHz at day 691 of 11.3$\pm$0.7\,mJy/beam.

Figure~\ref{contourplusopticalimage} shows
the full extent of the disk of M51 as far out as 16\,kpc away from the galactic centre. {\em This is the largest
extent of M51 detected so far in radio continuum.}\
The emission is not increasing uniformly with radius. An extension of the disk to the northeast is possibly caused by an outflow generated by the interaction with the companion.

\section{Spectral properties of M51 \label{section4}}

\subsection{Integrated radio continuum spectrum \label{sec:intspectrum}}

To calculate the flux density of M51 we integrated the total emission in concentric rings whilst taking into
account the major axis and inclination of the galaxy's plane projected on the sky.
We obtained an integrated flux density measurement of $S_{151}=8.1 \pm 0.6$\,Jy out to a
radius of 16\,kpc. The star forming region ($r < 10$\,kpc) contains 6.98\,Jy and the extended
disk ($r \textgreater 10$\,kpc) contains approximately 1.1\,Jy of flux density.

We used other measurements from higher frequencies, given in Table~\ref{integratedflux}, to compute the
spectral index of M51. As we have used the RBC flux scale for scaling our LOFAR data, measurements at other
frequencies must be rescaled.
Fortunately, at frequencies $\gtrsim$ 300\,MHz, the RBC scale is in agreement with KPW scale \citep{1969ApJ...157....1K}
and therefore we can use the conversion factors from the Baars scale (Table 7 in \citet{1977A&A....61...99B}).

Comparing to previous flux measurements, the LOFAR flux density agrees very well with other lower frequency data as well
as to the higher frequency data. The 4.86 and 8.46\,GHz values by \citet{2011MNRAS.412.2396F} have significant
uncertainties due to addition of single dish data to interferometry data.
The integrated flux of M51 from the 6C \citep{1988MNRAS.234..919H} ($6.90 \pm 0.69$\,Jy ) and 7C \citep{1996MNRAS.282..779W} ($6.48 \pm 0.65$\,Jy) surveys are very comparable the flux
of the star forming region of M51 (6.98\,Jy).

\begin{figure}
\resizebox{\hsize}{!}{\includegraphics{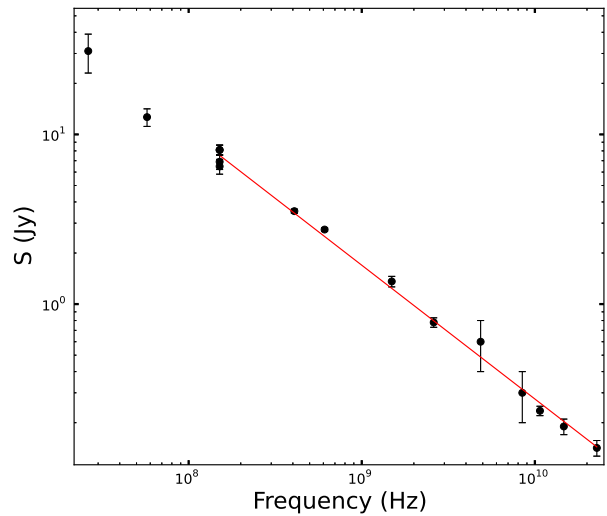}}
\caption{The integrated flux density of M51 with a power law fit of slope $\alpha$ = -0.79 $\pm$ 0.02. The integrated flux values of 11 $\pm$ 1.5\,Jy from \citet{1990ApJ...352...30I} at 57.5\,MHz and 31 $\pm$ 8 Jy at 26.3\,MHz \citep{1975AJ.....80..931V} are also plotted.}
\label{integratedfluxpic}
\end{figure}

We fit a single power law with a spectral index of $\alpha$ = $-0.79$
$\pm$ 0.02 (Fig.~\ref{integratedfluxpic}).
There is no indication of flattening of the integrated spectrum for M51
at low frequencies down to 151\,MHz due to thermal
absorption or ionisation losses.
The flattening towards 57.5\,MHz is uncertain, especially when compared to the flux measurement of 26.3\,MHz.
The spectral index between 26.3\,MHz and our LOFAR flux is approximately $\alpha$ = $-0.76$, showing no
significant flattening
of the spectrum down to 26.3\,MHz. This shows the importance of upcoming LOFAR LBA data.
Towards high frequencies no flattening by
thermal emission (with a spectral index of $-0.1$) is observed. The
average thermal fraction at 4.86\,GHz is about 25\%
from the observed spectral index and assuming a constant
synchrotron spectral index across the galaxy disk \citep{2011MNRAS.412.2396F},
which is an overestimate if the synchrotron spectral index is flatter in
the spiral arms \citep{2007A&A...475..133T},
and about 12\% when integrating the luminosity function of the {\sc H\,ii} regions
\citep{1988A&A...195...38V}.
Assuming 16\% at 4.86\,GHz, this should increase to 36\% at 22.8\,GHz
and flatten the spectral index
between 4.86 and 22.8\,GHz to $-0.63$,
which is clearly not observed. The flattening is possibly
cancelled by a steepening of the nonthermal synchrotron spectrum by
synchrotron and inverse Compton losses (see discussion in
Sect.~\ref{section9}).
The steepening of the synchrotron spectrum should become significant
beyond 2\,GHz, where the
average thermal fraction exceeds 10\%.
Another possibility is a lack of detected flux density at the highest
frequencies
listed in Table~\ref{integratedflux} where the signal-to-noise
ratios are lowest. This makes future observations of galaxies at frequencies above 20 GHz vital.

\begin{table}
\caption{Integrated flux densities for M51 from literature, rescaled to the RBC scale,
which are used for the fit in Fig.~\ref{integratedfluxpic}.}
\centering
\begin{tabular}{l c c}
\hline\hline
$\nu$ (GHz) & Flux density (Jy) & Ref \\
\hline
22.8 & 0.147 $\pm$ 0.016 & 1  \\
14.7 & 0.197 $\pm$ 0.021 & 1 \\
10.7 & 0.241 $\pm$ 0.014 & 2 \\
8.46 & 0.308 $\pm$ 0.103 & 3 \\
4.86 & 0.604 $\pm$ 0.201 & 4 \\
2.604 & 0.771 $\pm$ 0.049 & 1 \\
1.49 & 1.36 $\pm$ 0.09 & 4 \\
0.61 & 2.63 $\pm$ 0.06 & 5 \\
0.408 & 3.5 $\pm$ 0.1 & 6 \\
0.15 & 6.9 $\pm$ 0.69 & 7 \\
0.15 & 6.48 $\pm$ 0.65 & 8 \\
0.15 & 8.1 $\pm$ 0.6 & this work \\
\hline
\end{tabular}
\label{integratedflux}
\tablebib{(1) \citet{1984A&A...135..213K}; (2) \citet{1981A&A....94...29K}; (3) \citet{2011AJ....141...41D-1}; (4) \citet{2011MNRAS.412.2396F}; (5) \citet{1977A&A....54..703S}; (6) \citet{1980A&AS...41..329G}; (7) \citet{1988MNRAS.234..919H}; (8) \citet{1996MNRAS.282..779W}}
\end{table}

\subsection{Spectral index map of M51 \label{sec:spectrum}}

\begin{figure*}
\centering
\includegraphics[width=18cm]{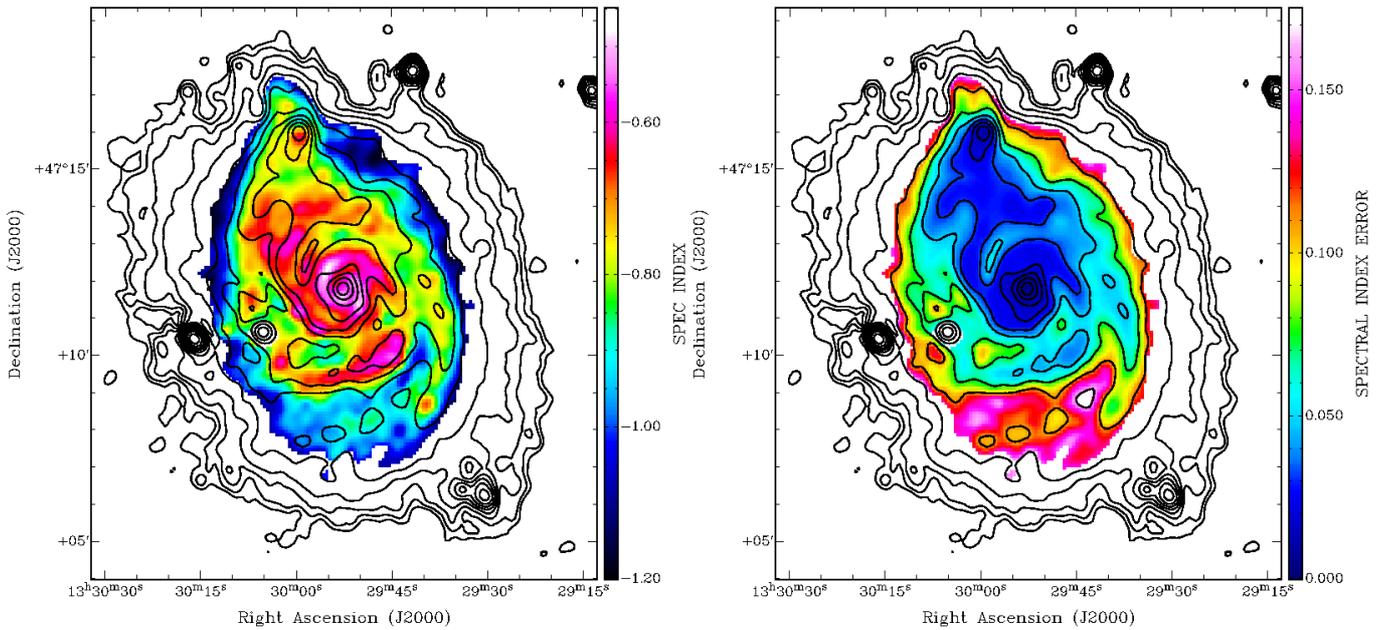}
\caption{Spectral index image (left) and spectral index error image (right) made from the 1.4\,GHz and 151\,MHz
images at
20\arcsec\ resolution. The 151\,MHz image is overlayed with the same contour levels as Fig.~\ref{contourplusopticalimage}.}
\label{specindexmap}
\end{figure*}

A spectral index image was created from the VLA image of M51 at 1.4\,GHz and the 151\,MHz LOFAR image. Both images were made with the same uv distance (namely 0.1 to 16 k$\lambda$). The 1.4\,GHz image was
convolved to 20\arcsec resolution and
placed onto the same grid as the 151\,MHz LOFAR image. Nearby point sources were either subtracted by fitting a Gaussian component and then
subtracting or (for sources within M51) manually blanked out. Only pixels that are above $5\sigma$ level in both images were used.
The spectral index was computed pixel by pixel and is shown in Fig.~\ref{specindexmap} along with the uncertainty image.

\begin{figure}
\resizebox{\hsize}{!}{\includegraphics{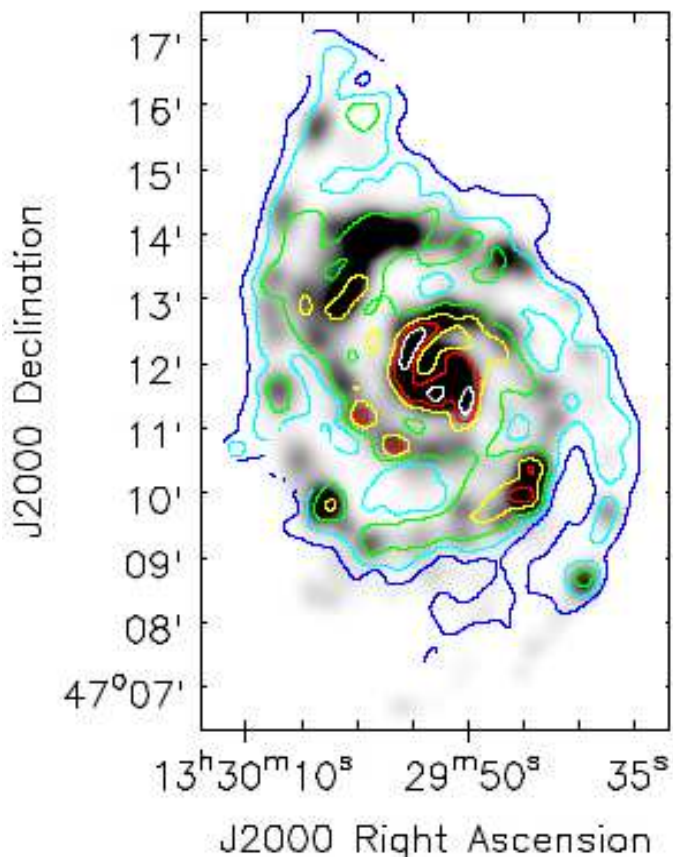}}
\caption{{\sc H\,ii} from \citet{2001AJ....122.3017S} smoothed to 20\arcsec\ resolution, with the map
of spectral indices between 1.4\,GHz and 151\,MHz
overlayed with contours. The white, red, yellow, green, cyan and blues contours show $-$0.5, $-$0.55, $-$0.6, $-$0.7, $-$0.8, $-$0.9, respectively.}
\label{HIIspecindex}
\end{figure}

A contrast between spiral arm and interarm regions is obvious in the spectral index image, indicating energy losses of
the CREs as they propagate from the star forming regions into the interarm regions.
According to the {\sc H\,ii} overlay (Fig.~\ref{HIIspecindex}), regions of strong star formation
in the spiral arms show flat spectral indices. Regions of high star formation are bulging into the interarm region,
specifically in the inner southeastern region where several {\sc H\,ii} regions exist in the interarm regions \citep{2011ApJ...735...75L}.

\subsection{Spectral radial profile of M51}

\begin{figure}
\resizebox{\hsize}{!}{\includegraphics{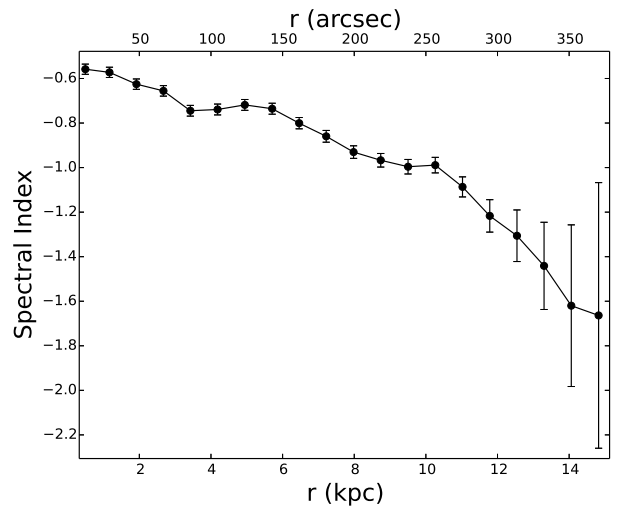}}
\caption{Radial profile of the spectral index between 1.4\,GHz and 151\,MHz.}
\label{radialspectral}
\end{figure}

Figure~\ref{radialspectral} shows the radial spectral index profile, created from the mean flux densities at 1.4\,GHz and
151\,MHz in concentric rings in the galaxy's plane and then computing the spectral index. The depressions
around r $\sim$ 2\,kpc and a larger dip at r $\sim$ 3.5\,kpc are associated with the interarm regions of M51.
The spectral index rises again as the spiral arm region becomes more dominant.
In the range 4\,kpc $<$ r $\leq$ 10\,kpc the spectral index gently decreases with a slight upturn at around
r $\sim$ 10\,kpc, signifying the location of the companion galaxy NGC5195.
At r $\textgreater$ 10\,kpc the spectral index drops rapidly, indicating that CREs
are significantly older in the outer disk compared to the inner star forming regions of the galaxy.

The extended disk shows  $\alpha \leq -0.9$, indicating energy losses through
synchrotron cooling and inverse Compton losses. Fig.~\ref{radialspectral} reveals a sharp decrease in
$\alpha$ beyond 10\,kpc, where the star formation rate is one
order of magnitude smaller than in the central disk \citep{2006ApJ...651L.101T}, resulting in hardly any
fresh injection of CREs.

The outer disk of M51 towards the companion galaxy NGC5195 behaves differently. The region between the companion
and the northern arm of M51 has a spectral index of $\sim -0.8$.
The northern spiral arm continues far beyond the optical arm to the left of the companion.
The arc shaped structure located just below NGC5195 seen in radio continuum \citep{2011AJ....141...41D-1} and in {\sc H\,ii} \citep{1998ApJ...506..135G} has a spectral index of $\sim -0.74$.
These results indicate that CREs are injected locally in the region around the companion galaxy, for example by shock
fronts generated by the interaction.

\begin{figure}
\resizebox{\hsize}{!}{\includegraphics{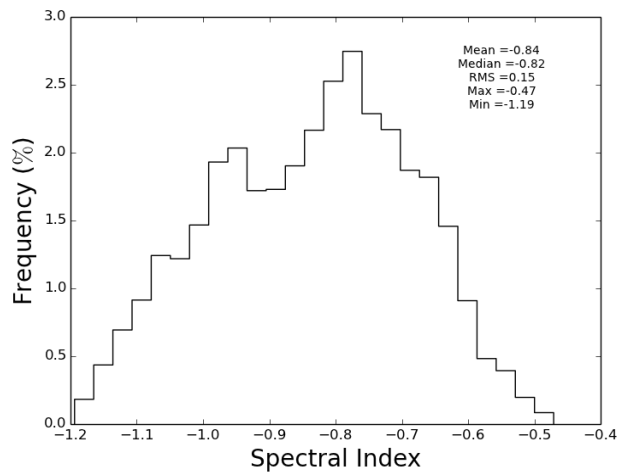}} 
\caption{Histogram of spectral index distribution grouped into 25 bins.}
\label{spectralhistogram}
\end{figure}

In contrast, the outer disk in the south has a spectral index of about
-1.0, causing a secondary peak in the spectral index histogram shown in Fig.~\ref{spectralhistogram}, which can be explained by CRE
diffusion from the spiral arms. The median value ($-0.82$) of the spectral index found in Fig.~\ref{spectralhistogram} agrees quite well with the integrated spectrum of M51 with
$\alpha$ = $-0.79$ $\pm$ 0.02.

\subsection{Evidence of free-free absorption in M51}

Primary CREs are believed to be injected into galaxies via supernova remnants (SNR) with a power-law spectra $Q(E) \propto E^{p}$ where the spectral index
$p$ of the energy spectrum is related to the spectral index $\alpha$ of the radio synchrotron emission
via $\alpha = (p-1)/2$ and $Q(E)$ is the CR source term.
Models of diffusive shock acceleration in strong shocks predict $p \sim -2$ \citep{1978MNRAS.182..147B}, consistent
with observational results based on Galactic SNR from \citet{2006A&A...457.1081K} and \citet{2009BASI...37...45G}
suggesting a mean nonthermal spectral index of $\alpha \sim -0.5$, which is indicative of a mature adiabatically
expanding SNRs \citep{2006A&A...457.1081K}. The energy spectrum of CRE, in the ISM is steepened to $p \sim -2.6 $
due to energy dependent diffusion \citep{1979ApJ...228..293J}, yielding a nonthermal radio spectral index of
$\alpha \sim -0.8$.

We observe a spectral index of $-0.47\geq \alpha \geq-0.52$ in the central region and inner spiral arms of M51 (see
Fig.~\ref{spectralhistogram} and Fig.~\ref{HIIspecindex}), which is flatter than expected from CRE acceleration models.
Using the following equation we can calculate the thermal fraction at 151\,MHz:

\begin{equation}
f_\mathrm{th}= \frac{q^{\alpha} - q^{\alpha_{syn}}}{q^{\alpha_{th}} - q^{\alpha_{syn}}}\, ,
\label{equ:thermalfraction}
\end{equation}
where $q$ is the ratio of the two frequencies (1400\,MHz and 151\,MHz), $\alpha$ is the mean
spectral index observed between the two frequencies, $\alpha_{syn}$ is the assumed synchrotron spectral index
and $\alpha_{th}$ is the assumed thermal spectral index.
If the flattening is caused entirely by free-free emission, $f_\mathrm{th}=$20--30\% at 151\,MHz would be needed,
assuming spectral indices
of $-0.8$ for the synchrotron and $-0.1$ for the thermal emission. This is unlikely at such a low-frequency, because a similar thermal fraction has been found at 4.86\,GHz \citep{2011MNRAS.412.2396F}, while the thermal fraction must strongly decrease towards lower frequencies.

\citet{2001AJ....122.3017S} found 1373 {\sc H\,ii} emission regions exist within the central region with sizes up
to 100\,pc, and therefore {\em thermal free-free absorption}\ is expected.
Recently, \citet{2013A&A...555A..23A} observed a spectral flattening due to free-free absorption at 350\,MHz in the
core region of M82 where intense star formation is known to be occurring. Future observations of M82 at lower frequencies,
specifically with LOFAR LBA, may show that this region becomes opaque.

\section{Total magnetic field of M51 \label{section5}}

\begin{figure}
\resizebox{\hsize}{!}{\includegraphics{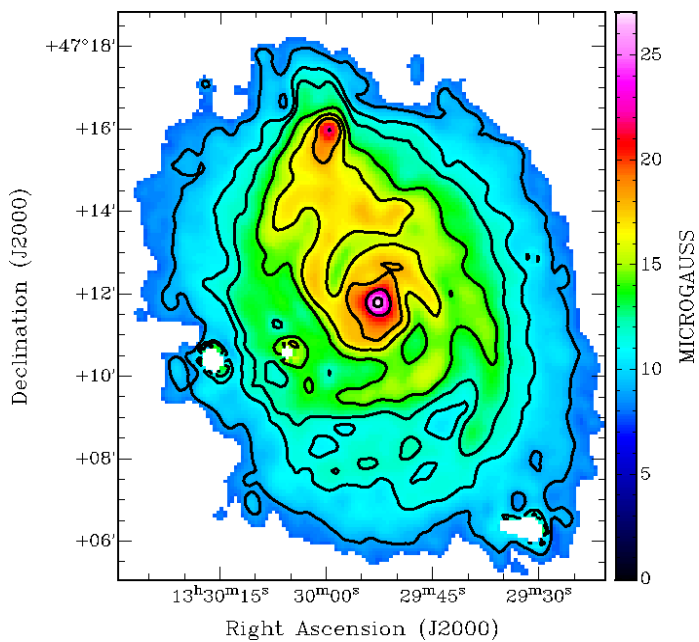}}
\caption{The total magnetic field of M51 in $\mu$G reaching up to $r \sim$ 10\,kpc, determined by assuming
energy equipartition. Contours are in levels of 9, 11, 12, 14, 16, 18, 22 and 26\,$\mu$G.}
\label{Bfieldpic}
\end{figure}

The total magnetic field strength of M51 can be determined from the synchrotron emission by assuming
equipartition between the energy densities of cosmic rays and magnetic field, using
the revised formula of \citet{2005AN....326..414B}. The total magnetic field strength scales with the synchrotron intensity $I_\mathrm{syn}$ as:

\begin{equation}
 B_\mathrm{tot,\perp} \propto I_\mathrm{syn}^{\,\,1/(3-\alpha)} \, ,
\end{equation}

\noindent where $B_{\mathrm{tot},\perp}$ is the strength of the total field perpendicular to the line of sight.
Further assumptions are required on the synchrotron spectral index of $\alpha=-0.8$ and the effective pathlength
through the source of 1000\,pc/cos({\it i}) = 1064\,pc.
We also assumed that the polarised emission emerges from ordered fields with all possible inclinations.
Here we assume a ratio of CR proton to electron number densities of $K_0=100$, which is a reasonable assumption
in the star forming regions in the disk \citep{1978MNRAS.182..443B}.
Large uncertainties for the pathlength and $K_0$ of a factor of 2 would effect the result only by 20$\%$. The effect of
adjusting $\alpha$ to between 0.7 and 0.9 produces an error of 5$\%$ in magnetic field strength.

\begin{figure}
\resizebox{\hsize}{!}{\includegraphics{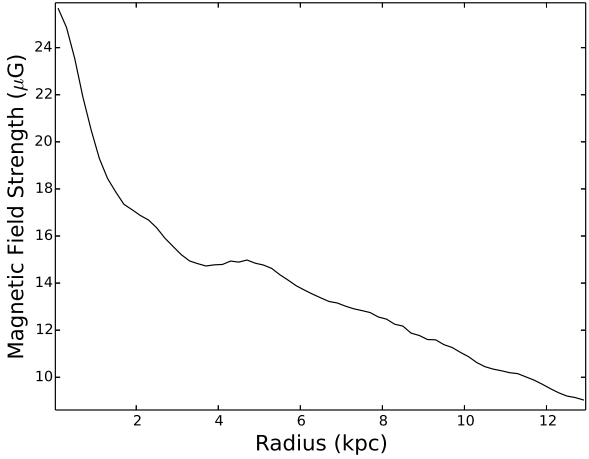}}
\caption{The radial profile of the total magnetic field of M51 in $\mu$G reaching out to 12\,kpc radius.}
\label{Bfieldpicradial}
\end{figure}

Using these assumptions, we created an image of the total magnetic field in M51 shown in Fig.~\ref{Bfieldpic}.
From this image we also created a radial profile of the total magnetic field strength, shown in Fig.~\ref{Bfieldpicradial}.
From these figures it is seen that the central region has a total magnetic field strength of between 20 and 30\,$\mu$G, the spiral arms of 10 to
20\,$\mu$G and the interarm
regions of between 10 and 15\,$\mu$G. At regions $r \sim$ 10\,kpc we observe field strengths of $\sim 10\,\mu$G.
These values are lower than the values found in \citet{2011MNRAS.412.2396F}, who performed a crude separation of
thermal and nonthermal synchrotron emission components and derived a synchrotron image to compute the magnetic field
strengths. The thermal component is much smaller at 151\,MHz than at 4.8\,GHz,
so we expect a smaller error in our estimates. These values are consistent with those modelled by \citet{2014arXiv1406.5717S}.

However, errors of the equipartition estimates will become significant in regions away from the CR sources,
especially in the outer disk where the observed factor $K$ will increase because energy losses of the CREs
are more severe than those of CR protons. Assuming equipartition underestimates the total magnetic field by a factor
of $(K/K_0)^{1/4}$ in these regions \citep{2005AN....326..414B}. Fortunately, energy losses are generally weaker at
low frequencies and therefore our equipartition estimate should be more accurate than higher frequency estimates.

\section{Cosmic ray propagation in M51 \label{section6}}

\subsection{Radial scale lengths of the total radio emission \label{section6a}}

The total extent of disk emission is insufficient as a measure of how far the disk extends as we are limited by the sensitivity of our observations. Therefore, we use the scale length $l$ which describes the emission along the disk as an exponential function, that is, $I_{\nu}  \propto \mathrm{exp}(-r/l)$, where $r$ is the galactocentric distance.

The radial profile of M51 was taken from the mean in concentric rings with the position angle of the major axis and the
inclination of the galaxy taken into account using the values from Table~\ref{physicalpara}. Surrounding background
point sources were removed by fitting Gaussians before measuring the radial profile.
Several background point sources located in the disk were blanked out.
The same was also done to the VLA (C+D arrays) 1.4\,GHz image, rescaled to the RBC flux scale from \citet{2011MNRAS.412.2396F}, which was placed onto the same grid and resolution as the 151\,MHz image.
The interarm region of M51 appears in the radial profile at 3.3\,kpc radius at both frequencies, but is less
prominent at 151\,MHz.

\begin{figure}
\resizebox{\hsize}{!}{\includegraphics{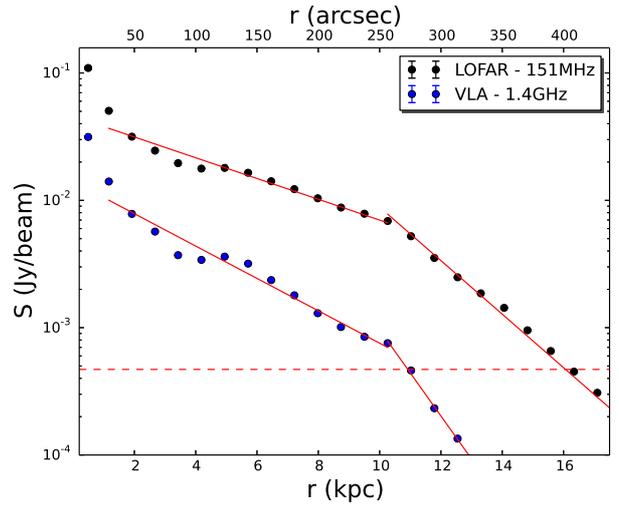}}
\caption{Radial profile of M51 at 151\,MHz and 1.4\,GHz. The horizontal red line shows the sensitivity limit
(3$\sigma$) of the 151\,MHz image. Errors are too small to be seen at this scale.}
\label{stokesiradial}
\end{figure}

A single exponential profile as fitted by \citet{2011AJ....141...41D-1} is not sufficient, because
a break occurs at around 10\,kpc at both frequencies (Fig.~\ref{stokesiradial}), just beyond the break in the
distribution of H$_2$ in M51 (see Fig.~4 in \citet{2007A&A...461..143S}). A break in the radial profile of M51 was also
detected in {\sc H\,i} at a radius of 9--10\,kpc from the galactic centre by \citet{2010AJ....140.1194B}
(see Table~\ref{scalelengthtable}).

It is very unlikely that the lack of short spacings can cause the break in the radial profile of M51. LOFAR has excellent inner uv coverage and is sensitive to structures over a half of a degree.
The same holds for the 1.4\,GHz VLA observations; in D configuration it is sensitive to structures up to 15\arcmin.

Two separate exponential functions were fitted for the inner and outer disk for both images:

\begin{equation}
  I(R) = \left\{
  \begin{array}{l l}
    I_\mathrm{0}  \, \mathrm{exp}(-r/l_\mathrm{inner}) & \quad \text{r $\le$ 10\,kpc}\\
    I_\mathrm{10} \, \mathrm{exp}(-r/l_\mathrm{outer}) & \quad \text{r $\ge$ 10\,kpc} \, .
  \end{array}
  \right.
\end{equation}

The radial profiles of the continuum emission at 151\,MHz and 1.4\,GHz with the fitted functions are shown in Fig.~\ref{stokesiradial}. The obtained scale lengths for the inner and outer parts of the galaxy are given in
Table~\ref{scalelengthtable}. The scale length at 1.4\,GHz in the inner disk derived by \citet{2011AJ....141...41D-1}
of 4.2$\pm$0.5\,kpc (scaled to the distance used in this paper) is somewhat consistent with the value derived here (3.4$\pm$0.2\,kpc).

\begin{table}
\caption{Scale lengths of the inner and outer disk of M51.}
\centering
\begin{tabular}{l c c}
\hline\hline
$\nu$ (MHz) & l$_\mathrm{inner}$ (kpc) & l$_\mathrm{outer}$ (kpc) \\
\hline
1400 & 3.4 $\pm$ 0.2 & 1.28 $\pm$ 0.02  \\
151 & 5.32 $\pm$ 0.4 & 2.06 $\pm$ 0.06 \\
{\sc H\,i}$^*$ & 5.5 & 2.1 $^*$\\
\hline
\multicolumn{3}{l}{$^*$ Estimated from Fig.~2 in \citet{2010AJ....140.1194B}.}
\end{tabular}
\label{scalelengthtable}
\end{table}

The scale length in the outer disk is 2.6~times smaller than the inner scale length at both frequencies. A similar
result has been obtained in M33 \citep{2007A&A...472..785T} where two exponential scale lengths were fitted.
The scale length for $r<$ 4\,kpc in M33 is twice as large as the scale length $r \textgreater$ 4\,kpc.

The scale length of radio synchrotron emission is determined by the radial profile of the total magnetic field,
distribution of CRE sources and CRE diffusion. The reason for the break is probably due to the break in
distribution of CRE sources, closely related to the H$_2$ distribution (see Fig.~4 in \citet{2007A&A...461..143S}), at about 6--7\,kpc, but shifted outwards by a few kpc due to CRE diffusion (see Sect.~\ref{section6b}).

The scale lengths in the inner and outer disk are larger at 151\,MHz than at 1.4\,GHz by a factor of $1.6\pm0.1$,
due to the fact that CRE energy loss processes are weaker at lower frequencies, the CRE lifetime is larger, and
hence the radial propagation length of CREs is larger, leading to a larger scale length.

We investigate two models of CRE propagation \citep{2013A&A...557A.129T}:

(A) {\em Diffusion}: The CRE propagation length depends on the
average perpendicular and parallel diffusion coefficient, $D$, and the
CRE lifetime, $\tau_\mathrm{CRE}$, as $l_\mathrm{dif} \propto (D\,\tau_\mathrm{CRE})^{1/2}$.
$D$ may depend weakly on the energy $E$ of the electrons for $E>4$\,GeV
\citep{1990A&A...233...96E}.
In the strong field of M51 ($B_\mathrm{tot} \simeq 15\,\mu$G), we trace electrons of 2.4 and 0.8\,GeV at
1400 and 151\,MHz, respectively, so that the energy dependence of $D$ can be neglected here.
As the lifetime of CRE is limited by synchrotron losses (Sect.~\ref{section6b}, Eq.~\ref{eq:sync}),
$\tau_\mathrm{CRE} =  \tau_\mathrm{syn} \propto B_\mathrm{tot}^{\,\,-2} \, E^{-1} \propto B_\mathrm{tot}^{\,\,-3/2} \, \nu^{\,\,-1/2}$,
where $\nu$ is the observation frequency. Insertion into the above relation gives:
\begin{equation}
l_\mathrm{dif} \propto B_\mathrm{tot}^{\,\,-3/4} \, \nu^{\,\,-1/4} \, .
\label{eq:diffusion}
\end{equation}
\noindent With a frequency ratio of 9.27, the ratio of propagation lengths is 1/1.74.

(B) {\em Streaming}: The CRE propagation length along the magnetic field in case of the streaming instability \citep{1969ApJ...156..445K}
depends on Alfv\'en velocity $v_\mathrm{A} \propto B_\mathrm{tot}$ and CRE synchrotron lifetime $\tau_\mathrm{syn}$:
\begin{equation}
l_\mathrm{stream} = v_\mathrm{A}\,\tau_\mathrm{CRE} \propto B_\mathrm{tot}^{\,\,-1/2} \, \nu^{\,\,-1/2} \, .
\label{eq:stream}
\end{equation}
\noindent With a frequency ratio of 9.27, the ratio of propagation lengths is 1/3.04.

The ratios of scale lengths observed at 1400 and 151\,MHz  of about 1/1.6 (Table~\ref{scalelengthtable})
agree with the diffusion model.

Both models are rather simplistic and should only be used to give tendencies about the intrinsic distribution
and the magnitude of the CRE propagation length. A more detailed analysis needs a proper description of CRE sources,
magnetic field distribution, CRE loss processes and CRE propagation mechanisms.

\subsection{CRE diffusion coefficients in M51 \label{section6b}}

CREs lose their energies via a number of different processes such as synchrotron radiation, inverse
Compton (IC) radiation, non-thermal bremsstrahlung, ionisation,
and adiabatic expansion. Out of these processes, synchrotron and inverse Compton losses have the same
dependence on particle energy and therefore are difficult to distinguish from the radio spectrum alone.
 Inverse Compton loss in the galaxy's radiation field is
generally smaller than synchrotron loss in galaxies \citep{2014AJ....147..103H}
and is neglected in the following.

\begin{figure*}
\centering
\includegraphics[width=18cm]{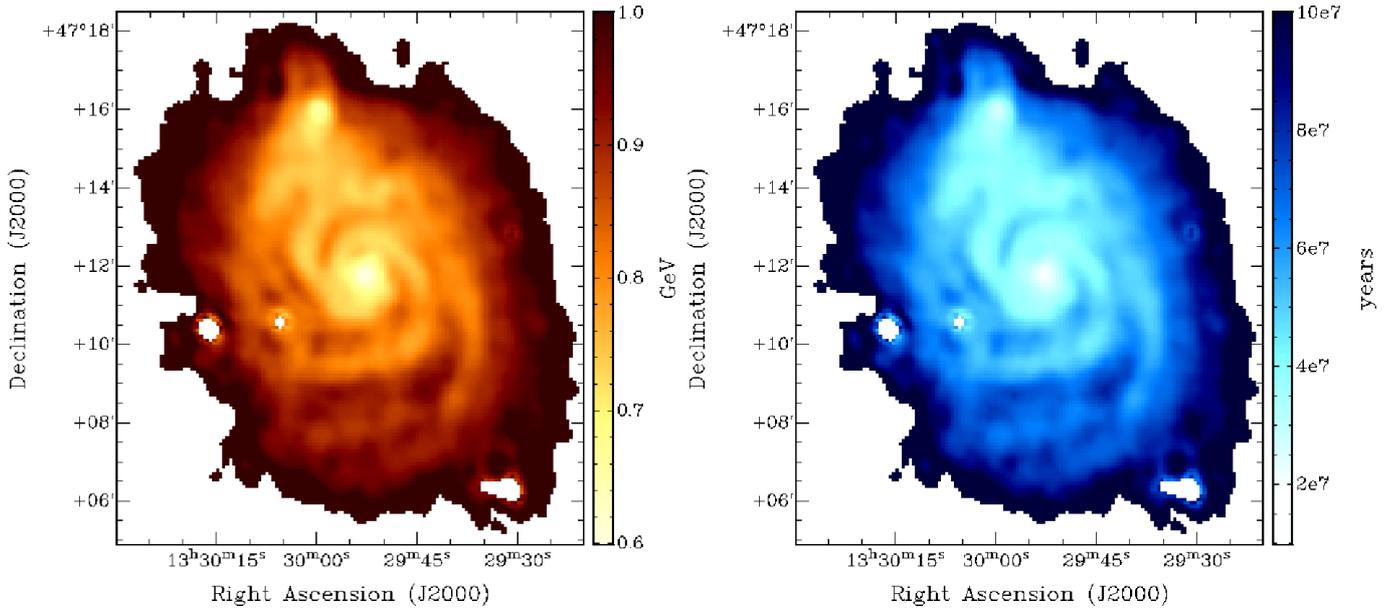}
\caption{Images of CRE energy (left) and the synchrotron lifetime (right) in M51.}
\label{energycosmicray}
\end{figure*}

We use the equations from \citet{1990A&A...239..424P} to calculate the CRE energy $E$ and the synchrotron lifetime throughout M51 (Fig.~\ref{energycosmicray}):

\begin{equation}
\left(\frac{E}{\rm{GeV}}\right) = \left(\frac{\nu}{16.1\,\rm{MHz}}\right)^{\frac{1}{2}} \, \left(\frac{B_\mathrm{tot,\perp}}{\rm{\mu G}}\right)^{-\frac{1}{2}}
\label{equationenergycosmicray}
\end{equation}
and
\begin{equation}
\left(\frac{\tau_\mathrm{syn}}{\rm{yr}}\right) = 8.352\times10^9 \,\, \left(\frac{E}{\rm{GeV}}\right)^{-1} \, \left(\frac{B_\mathrm{tot}}{\rm{\mu G}}\right)^{-2} \, .
\label{eq:sync}
\end{equation}

 If we take $l_\mathrm{dif}=$ 1.45\,kpc from Sect.~\ref{section7} as the typical distance that a
CRE travels from its origin, the diffusion coefficient D of the electrons can be estimated by
(see e.g. \cite{2014arXiv1406.6488I}):
\begin{equation}
D = \frac{l_\mathrm{dif}^2}{4  \, \tau_\mathrm{syn}} \, .
\label{diffusionequation}
\end{equation}

With an average synchrotron lifetime of $\simeq 4.8 \times 10^{7}$ years across the star forming region of the galaxy, we estimate
$D \simeq 3.3 \times 10^{27}$\,cm$^{2}$\,s$^{-1}$.
This value is smaller than that of \citet{2011MNRAS.412.2396F} who disregarded the factor of 4 in
Eq.~\ref{diffusionequation} and assumed $l_\mathrm{dif}=$ 1\,kpc at 5\,GHz which seems too high
compared to our results. \citet{2013A&A...557A.129T} estimated $l_\mathrm{dif}>$ 0.5\,kpc in M51
at 1.4\,GHz where $\tau_\mathrm{syn} \simeq 1.5 \times 10^{7}$ years, which yields 
$D > 1.3 \times 10^{27}$\,cm$^{2}$\,s$^{-1}$, consistent with our result. 
On the other hand, the values of $D \sim (2-4) \times 10^{28}$\,cm$^{2}$\,s$^{-1}$ used in the 
models by \citet{1998ApJ...493..694M} for the Milky Way are higher.
It is important to note that the diffusion coefficient may vary with magnetic field strength and
degree of field order. As M51 has a stronger and possibly more turbulent magnetic field than the Milky Way,
the diffusion coefficient in M51 could be lower.

Figure~\ref{energycosmicray} shows us that at the edge of the extended disk $r \sim$ 10\,kpc, the synchrotron lifetime for CREs is found to be approximately around
$10^{8}$\,yr $< \tau_\mathrm{syn} \leq 2 \times 10^{8}$\,yr.
Using a diffusion coefficient of $\sim 3.3 \times 10^{27}$\,cm$^{2}$\,s$^{-1}$
and Eq.~\ref{diffusionequation},
the electrons at $r \sim$ 10\,kpc are able to travel up to 2.1--3.0\,kpc, neglecting other energy losses.
This takes us out to the edge of the extended disk detected with LOFAR.

\subsection{Diffusion of CREs into the interarm regions \label{section6c}}

According to observations of several spiral galaxies such as NGC6946 with the GMRT at 333\,MHz, the arm and
interarm regions are indiscernible \citep{2012MNRAS.419.1136B}, while in the case of M51 at 151\,MHz the arm
and interarm regions are clearly separated. This is unexpected because at 151\,MHz emission emerges from
an even older population of CREs that should diffuse further away from star forming regions in the spiral arms.

\begin{figure*}
\centering
\includegraphics[width=18cm]{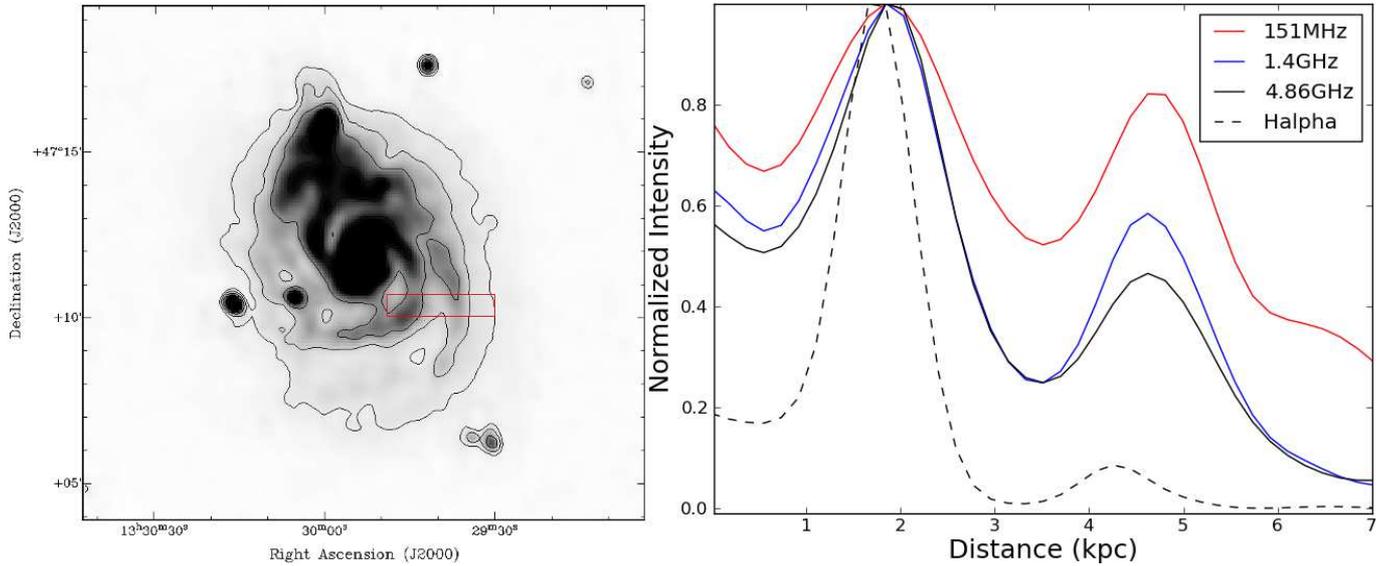}
\caption{Intensity profile showing the arm and interarm contrasts of H$\alpha$ and three different radio frequencies along a slice through M51. The location of this slice is shown in the image to the right. The peak flux densities from each image are normalised to 1.}
\label{diffusionplot}
\end{figure*}

To quantify our result, we used a H$\alpha$ image \citep{1998ApJ...506..135G}
and the 4.86\,GHz and 1.4\,GHz VLA images from \citet{2011MNRAS.412.2396F}
in addition to our LOFAR 151\,MHz image. All images were smoothed to a common resolution of 20\arcsec\ and
transformed to the same grid.
Fig.~\ref{diffusionplot} shows a slice of M51 for all 4 images through a region where the arm--interarm contrast could be best studied. The slice is taken at a fixed declination of
+47$^\circ$ 10$^\prime$ 23$^{\prime \prime}$ and extends from a right ascension of
13$^{\mathrm{h}}$ 29$^{\mathrm{m}}$ 30.1$^{\mathrm{s}}$ to 13$^{\mathrm{h}}$ 29$^{\mathrm{m}}$ 49.2$^{\mathrm{s}}$.

The 4.86\,GHz and 1.4\,GHz profiles are very similar, indicating that the CREs radiating at 1.4\,GHz,
which have about half of the energy, are not diffusing much further than at the 4.86\,GHz eletrons.
At 151\,MHz, tracing CREs with 3 times less energy compared to 1.4\,GHz, a much smoother gradient
indicates that there are considerably more CREs diffusing far into the interarm region.

While we see considerable CRE diffusion into the interarm region of M51, the question arises why we don't
observe CREs diffusing the same distances as in the 333\,MHz images of other galaxies by \citet{2012MNRAS.419.1136B}.
One reason could be a stronger and more turbulent magnetic field in the spiral arms of M51 compared to
galaxies like NGC6946. The shorter synchrotron lifetime (Eq.~\ref{eq:sync}) leads to a shorter propagation
length, making it harder to diffuse into the interarm regions.

\section{Wavelet cross-correlation and CRE propagation length \label{section7}}

Wavelet transforms have been applied to several images of galaxies, for example M33 ~\citep{2007A&A...466..509T}
and M51 ~\citep{2011AJ....141...41D-1}.
Wavelet transforms and cross-correlations allow us to separate the diffuse emission components from compact sources
and to compare the emission at different wavelengths which we shall perform in this section. It is also
useful when studying the radio--far-infrared (FIR) correlation at various spatial scales, especially
when separating the differences of the correlation between the arm and interarm regions (see for example  \citet{2011AJ....141...41D-1} and \citet{2012ApJ...756..141B}).

The wavelet coefficient is defined as:
\begin{equation}
 W(a,\mathbf{x}) = \frac{1}{a^{\kappa}} \int^{+\infty}_{-\infty} f(\mathbf{x^{\prime}}) \, \psi^{*} \,
 (\frac{\mathbf{x^{\prime}} - \mathbf{x}}{a})d \mathbf{x^{\prime}} \, ,
\label{waveletequation}
\end{equation}
\noindent where $\psi(\mathbf{x})$ is the analysing wavelet, $\mathbf{x} = (x,y)$, $f(\mathbf{x})$ is a two-dimensional
function which in this case is an image, $a$ and $\kappa$ are the scale and the normalisation parameters, respectively,
and finally the $^{*}$ symbol denotes the complex conjugate.

The Mexican Hat wavelet is a real isotropic wavelet with a minimal
number of oscillations and was used as we wish to have more independent points:
\begin{equation}
\psi(\rho) = (2-\rho^2) \,\, e^{-\rho^{2}/2} \, .
\end{equation}
Our LOFAR image at 151\,MHz, the VLA image at 1.4\,GHz and the 70\,$\mu$m image from the SPITZER-MIPS (Multiband Imaging
Photometer for Spitzer) \citep{2004ApJS..154...25R} (courtesy of F.~Tabatabaei) were used for this analysis.
FIR flux densities from normal galaxies are commonly taken to indicate the rate of recent star formation.
\citet{2004ApJS..154..259H} found that FIR emission at 24 and 70\,$\mu$m follows closely the structure of the ionised
gas in M33, indicating that it is heated mostly by hot ionising stars.

\begin{figure*}
\centering
\includegraphics[width=18cm]{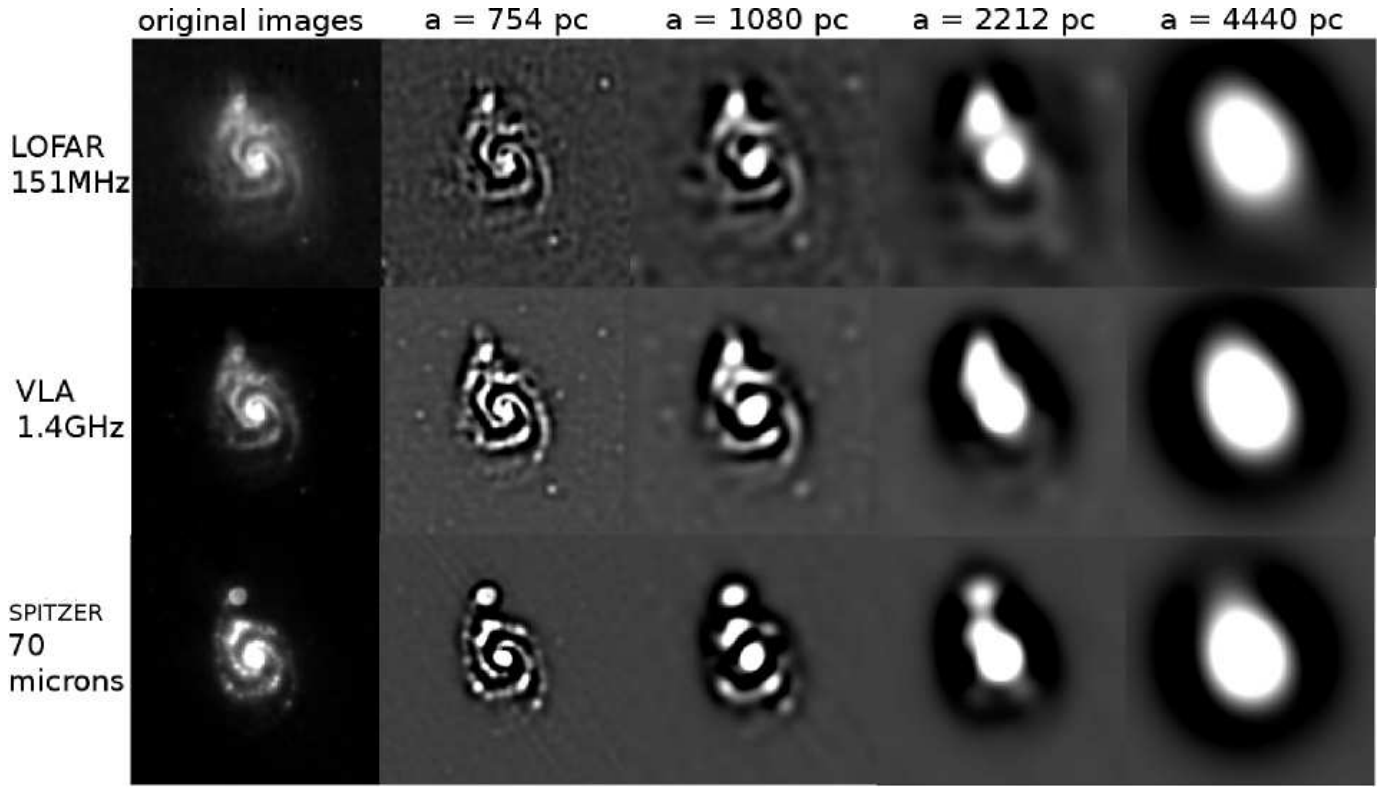} 
\caption{Wavelet decomposition of the three available datasets (151\,MHz, 1.4\,MHz and 70\,$\mu$m) at
four different spatial scales. The LOFAR 151\,MHz emission clearly extends farther out.}
\label{WAVELETPICS}
\end{figure*}

All images were made to have the same size, be on the same grid and were smoothed to a PSF with FWHM of 20\arcsec.
Bright background sources from the VLA and LOFAR images were subtracted from the images.
The images have been decomposed into 10 different scales $a$ with log spacing in order to compare
the morphology between the three different images (Fig.~\ref{WAVELETPICS}).

The individual {\sc H\,ii} regions within the spiral arms in the SPITZER and the VLA images coincide quite well at the smallest
scale ($a = 754$\,pc). However, these regions are not visible as well in the LOFAR image at the same scale.
The spiral arms are prominent at all three wavelengths at scales up to 1080\,pc. The spiral arms become indiscernible
at a scale of 2212\,pc, while in the LOFAR image, and to a smaller extent in the VLA image, the southern spiral arm
can be seen to extend much farther out.

At a scale of 4440\,pc, no discernible features can be seen at any of the wavelengths, only
the underlying diffuse disk with a gentle radial decrease in intensity. It should be noted
that the radial decrease is slower for M51 in the LOFAR dataset, consistent with Fig.~\ref{stokesiradial}.

The wavelet cross-correlation is a useful method to compare different images as a function of spatial scales
(\citet{2001MNRAS.327.1145F}, \citet{2011AJ....141...41D-1} and \citet{2007A&A...466..509T}).
Whilst normal cross-correlation analysis such as pixel to pixel correlation can be dominated by
bright extended regions or large scale structure, the wavelet cross-correlation allows
the analysis of a scale-dependent correlation between two images.

The cross-correlation coefficient at scale $a$ is defined as:

\begin{equation}
 r_{w}(a) = \frac{\int \int W_{1}(a,\mathbf{x})\,W_{2}^*(a,\mathbf{x}) d\mathbf{x}}{[M_{1}(a)\,M_2(a)]^{1/2}}
\end{equation}

where $M(a)$ is the wavelet spectrum:

\begin{equation}
M(a) = \int_{-\infty}^{+\infty} \int_{-\infty}^{+\infty} |W(a,\mathbf{x})|^{2}d\mathbf{x}
\end{equation}

\begin{figure}
\resizebox{\hsize}{!}{\includegraphics{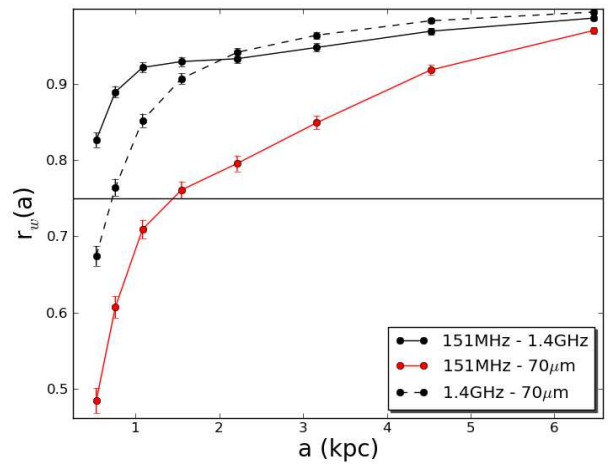}} 
\caption{Wavelet cross-correlation spectra between the three data sets. The correlation coefficient of $r_w = 0.75$
defining the break scale is shown by the solid line.}
\label{diffusionplot2}
\end{figure}

The value of $r_w$ can range between -1 (total anticorrelation) and +1 (total correlation).
Plotting $r_w$ against scale shows how well structures at different scales cross-correlate in intensity and location
between the two images (Fig.~\ref{diffusionplot2}).

The cross-correlation spectrum between the 1.4\,GHz and 151\,MHz images is extremely good for all
scales, except for scales less than 1\,kpc (Fig.~\ref{diffusionplot2}).
Overall, $r_\mathrm{w}$ for 151\,MHz -- 70\,$\mu$m is smaller compared to 1.4\,GHz -- 70\,$\mu$m.

The break scale at which $r_\mathrm{w} = 0.75$ is reached for a cross-correlation spectrum between radio
synchrotron and far-infrared emission is taken as a measure of the diffusion length $l_\mathrm{dif}$
\footnote{The absolute values of $l_\mathrm{dif}$ depend on the definition of the break scale.
$r_\mathrm{w} = 0.50$ was used by \citet{2013A&A...557A.129T} to define the break scale because
the correlations analysed in that paper were generally weaker than the ones in this paper.},
which depends on the diffusion coefficient and CRE lifetime as $l_\mathrm{dif} = 2\,(D\,\tau_\mathrm{CRE})^{0.5}$
(see Sect.~\ref{section6a}).

From Fig.~\ref{diffusionplot2} we derive that $l_\mathrm{dif}$ is approximately 720\,pc for the cross-correlation
between 1.4\,GHz and 70\,$\mu$m and about 1.45\,kpc between 151\,MHz and
70\,$\mu$m. This is interpreted as the lower-energy CREs propagating from the star forming regions have longer
lifetimes and travel further.

The ratio of propagation lengths of 2.0 is consistent with diffusive CRE propagation (Eq.~\ref{eq:diffusion}).

\citet{2011AJ....141...41D-1} (their Fig.~7) derived wavelet cross-correlation spectra only between the total
radio and 24\,$\mu$m data which cannot be directly compared with the results in this paper. Still, a trend
of $l_\mathrm{dif}$ increasing with decreasing frequency is also seen.

Our value for $l_\mathrm{dif}$ in M51 between 1.4\,GHz and 70\,$\mu$m is smaller than that for M31
($l_\mathrm{dif} = 3\pm1$\,kpc for our definition of the break scale) and also for NGC6946
($l_\mathrm{dif} = 5\pm1$\,kpc)
\citep{2013A&A...557A.129T}. This further supports our result from Sect.~\ref{section6c} that the typical
propagation length of CREs in M51 is small due to their short synchrotron lifetime in the strong field.
This also answers the question why the arm and interarm contrast is still seen in M51 at 151 MHz but not in NGC6946 at 333 MHz \citep{2012MNRAS.419.1136B}.

M51 and M31 have a similar degrees of magnetic field order, so that the relation between $l_\mathrm{dif}$ and the
degree of field order proposed by \citet{2013A&A...557A.129T} predicts similar values of $l_\mathrm{dif}$. However,
we find a value only half as large in M51. \citet{2013A&A...557A.129T} assumed that the CRE lifetime is given
by the energy-independent confinement time of about $3\times10^7$\,yr within the disk that is limited by escape loss,
while in M51 the CRE lifetime is limited by the shorter synchrotron lifetime (see Fig.~\ref{energycosmicray} right).

\section{Detection of polarised sources \label{section8}}

\subsection{Introduction}

Previously, very little has been done to study polarisation at frequencies below 200\,MHz, and polarisation
characteristics cannot be extrapolated
directly from higher frequencies. From previous observations at 350\,MHz \citep{2009A&A...494..611S}, one would optimistically expect to detect one polarised source with a polarisation degree of a few percent for every four square degrees on the sky \citep{2013ApJ...771..105B}. It is worth noting that for the observations of \citet{2009A&A...494..611S} at 350\,MHz, the resolution was
$2.7 \times 4.7$ arcmin with a noise of polarised intensity of 0.5\,mJy/beam.

Studies of polarisation at frequencies below 300\,MHz are now becoming possible, partly due to the construction
of LOFAR \citep{2013A&A...556A...2V} and the Murchison Widefield Array (MWA) \citep{2009IEEEP..97.1497L}, and to the introduction of new techniques to analyse polarisation,
in particular RM synthesis \citep{2005A&A...441.1217B}.
Recently, \citet{2013ApJ...771..105B} performed a 2400 square degree polarisation survey at 189\,MHz with the MWA
with a 7\arcmin\ beam. Out of a catalogue of 137 sources brighter than 4\,Jy in flux density, only one source
was detected in polarisation. Both beam depolarisation and internal Faraday dispersion could reduce the measured polarised
emission, therefore higher resolution observations are essential.
\citet{2013A&A...559A..27G}, whilst detecting diffuse polarisation in a nearby galaxy -- namely M31 -- for the first
time below 1\,GHz, also created a catalogue of 33 polarised background sources.
Extrapolation of these results to the frequencies of LOFAR would result in very few source detections and no
detections of polarised diffuse emission from the star forming disks of galaxies.
However, these observations are limited by the large angular resolution of 4\arcmin. It needs to be seen
if beam depolarisation plays a major role.
\citet{2013arXiv1309.4646F} applied RM Synthesis to data from the Giant Meterwave Radio Telescope (GMRT) at 610\,MHz
and found that M51 is depolarised to below the sensitivity limit.

LOFAR is the perfect instrument to measure Faraday depths with a high precision
and hence should allow us to detect weak magnetic fields and low electron densities that are unobservable
at higher frequencies. Importantly, LOFAR with its high sensitivity and high angular resolution should be able
to reduce depolarisation effects.

LOFAR has already produced polarisation results: \citet{2013A&A...558A..72I} were able to detect a faint and
morphologically complex polarised foreground of the highly polarised Fan region, in agreement with previous WSRT observations \citep{2009A&A...500..965B}.

\subsection{Applying RM Synthesis}

For each frequency channel, Stokes Q and U were imaged at 20\arcsec\ resolution with an area of 17.3 square degrees with CASA.
Therefore no primary beam correction was applied, which leads to errors.
However, using AWimager for a field of this
size and at this high resolution for every frequency channel would require much more processing time and power than
available. Each channel image was inspected and all channels with RFI contamination were discarded. This resulted
in 3774 final images.

Rotation measure (RM) synthesis was performed on these images using software written by the LOFAR Magnetism Key Science Project (MKSP).
The rms noise for the uncorrected primary beam image for polarised intensity was found to be
100\,$\mu$Jy/beam/rmsf. For Stokes U and Q, the rms noise was found to be 123\,$\mu$Jy/beam and 132\,$\mu$Jy/beam, respectively. The resulting Faraday cube at 20\arcsec\ resolution was then cleaned, using the RM Clean code of \citet{2009A&A...503..409H} for MIRIAD \citep{1995ASPC...77..433S}.
A maximum of 1000 iterations was used as well as a 1\,$\sigma$ cutoff level.

Usually, instrumental polarisation is located at a Faraday depth of 0\,rad\,m$^{-2}$. However, due to the
ionospheric RM correction that was performed, this instrumental polarisation is shifted
by the average RM correction applied to the data. Therefore, for this observation, the instrumental
polarisation is located at a Faraday depth of approximately $-$1\,rad\,m$^{-2}$.

The resolution in Faraday depth $\phi$ is given by the measured full width half maximum
of the RM spread function (RMSF) \citep{2005A&A...441.1217B}:
\begin{equation}
 \phi = \frac{2\,\sqrt{3}}{\Delta\lambda^2}
\end{equation}
where $\Delta\lambda^2$ is the width of the observed $\lambda^2$ distribution.

A minimum frequency of 115.9\,MHz and a maximum of 176\,MHz gives us $\phi = 0.91$\,rad\,m$^{-2}$.
The largest detectable structure in Faraday spectrum \citep{2005A&A...441.1217B} is given by:
\begin{equation}
  \phi_{max} = \frac{\pi}{\lambda_{min}^2}
\end{equation}
For our observations $\phi_{max} = 1.085$\,rad\,m$^{-2}$. Any larger structure will be depolarised.

The error in Faraday depth is found by the following expression:
\begin{equation}
 \Delta \phi = \frac{\phi}{2\,S/N}
\end{equation}

\noindent where $\phi$ is the FWHM of the RMSF and $S/N$ is the signal-to-noise ratio of the peak of the
component in the Faraday spectrum.

Following the foreground RM measured by \citet{2011MNRAS.412.2396F}, \citet{2009A&A...503..409H} and \citet{1992A&A...265..417H}, we would expect the polarised signal of M51 to be centred on $+ 9$\,rad\,m$^{-2}$.
At low frequencies, the polarised signal of M51 should be centred
around this value, with a dispersion smaller than that in Fig.10 of \citet{2011MNRAS.412.2396F} because only the halo and the foreground contribute to
the dispersion.

{\em No obvious diffuse polarisation from M51 could be detected} by looking into the FD-cubes. An integration over 25\,rad m$^{-2}$ wide parts in Faraday-space, for the area of M51, that is above the 1\,mJy contour in the total power image, delivered only noise with a standard deviation of $\sim$100$\mu$Jy. The same integration was carried out for four regions with the same size close to M51. Those delivered a slightly lower standard deviation of 80$\mu$Jy. For all integrations the region from $-$5\,rad m$^{-2}$ to +5\,rad m$^{-2}$ was excluded to not include the instrumental peak around $-$1\,rad m$^{-2}$. We attribute the slightly higher standard deviation in the M51 field to the remnants of the sidelobes of the instrumental peaks. Therefore we set a 5$\sigma$ detection limit for the integrated polarised intensity of M51 to 0.5\,mJy which corresponds to a polarisation degree of $0.006\%$.

This is not surprising in view of the small $\phi_{max}$ and is in line with
\citet{2013arXiv1309.4646F} who were unable to detect M51 in polarisation at 610\,MHz with the GMRT.

\subsection{Extragalactic polarised sources}

Taking five times the rms  noise of 100\,$\mu$Jy/beam/rmsf as the detection threshold and assuming an average degree
of polarisation of 1\%, all sources with flux densities above 50\,mJy/beam were checked for polarisation,
with six detections of polarised background extragalactic radio galaxies in the field.
Two of these sources are partly resolved, with polarisation detected in the lobes. Three sources were found to be outside the FWHM of the station beam of 3.6 degrees.
This results in a mean detection rate of one polarised source for every 1.7 square degrees.
As these sources are several degrees from the phase centre, the primary beam becomes important. To apply the primary beam,  the fluxes of these polarised sources were scaled by comparing a primary beam corrected image created from AWimager and the final CASA image. The difference for each source was found between the two images and applied. This corrected flux is shown in the following Faraday spectra and Table~\ref{polarisedtable}.

Where data are available from higher frequencies, notably from \citet{2009ApJ...702.1230T}, we can determine the
depolarisation ratio as defined in \citet{2007A&A...470..539B}:
\begin{equation}
DP(151,1400) = (PI_{151}/PI_{1400})\, (\nu_{1400}/\nu_{151})^{\alpha}
\end{equation}
\noindent where $\alpha$ is the synchrotron spectral index and $\nu_i$ is the respective frequency.
A value of 1 means no additional depolarisation at 151\,MHz with respect to 1400\,MHz.

In the following, we will only address the resolved sources briefly.
A summary of all detections are given in Table~\ref{polarisedtable}. The Faraday spectra of the remaining sources can be seen in Fig.~\ref{all4faradayspec}.
\subsubsection{J133920+464115}

Strong polarisation was detected in this single lobe radio galaxy B3 1337+469, solely from the lobe itself (J133920+464115) that contains two spatially separated components observable in both total and polarised intensity.
The northern component, which is brighter, is detected at a Faraday depth of $+20.5\pm0.1$\,rad\,m$^{-2}$, while the southern
component has $+20.75\pm0.1$\,rad\,m$^{-2}$.

\begin{figure*}
\centering
\includegraphics[width=16cm]{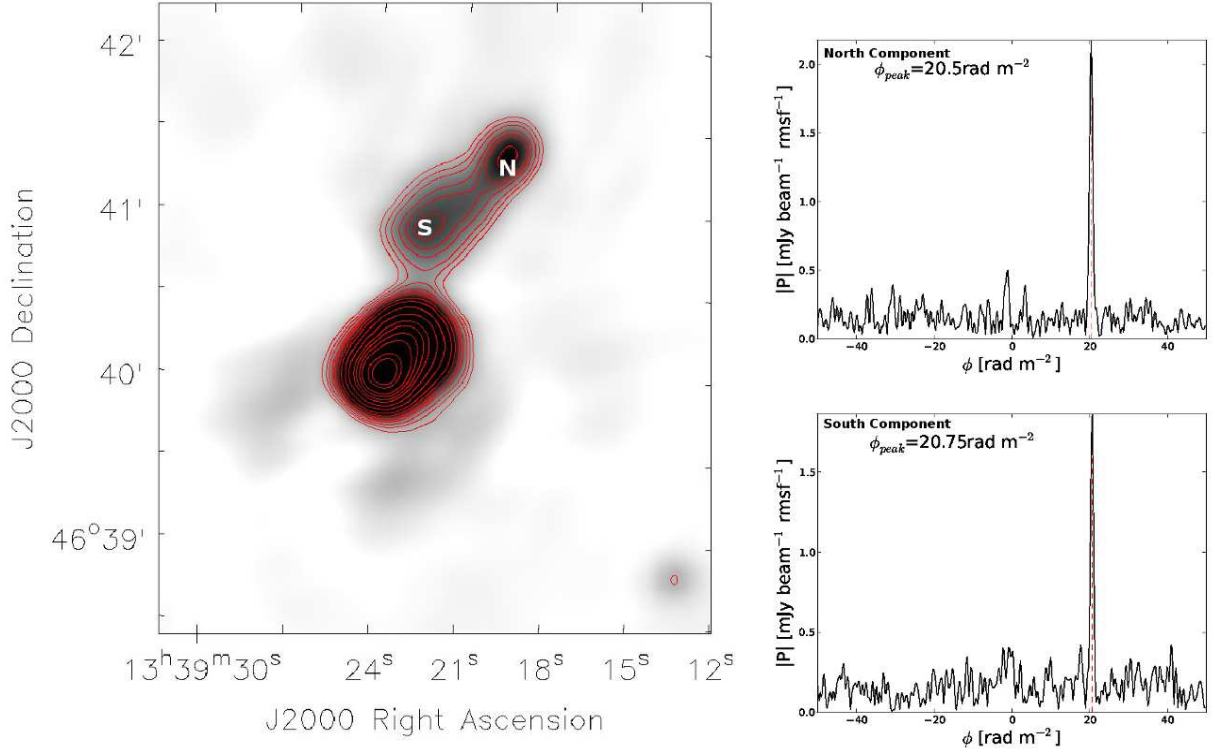}
\caption{The radio lobe J133920+464115 seen in total intensity (left) with the two marked spatial components
and the corresponding Faraday spectra (right). $rmsf$ is the half-power width of the RM spread function.}
\label{faradayspectra1}
\end{figure*}

The Faraday spectra and total intensity image for this source are shown in Fig.~\ref{faradayspectra1}.

This source is only partially resolved in the NVSS and from \citet{2009ApJ...702.1230T} it is polarised in the core with a RM of $22.5 \pm3.9$ \,rad\,m$^{-2}$
and a polarisation degree of $3.66 \pm 0.07 \%$. This is completely depolarised at 151\,MHz.
At 1.4GHz, the jet component has a RM of $5.5 \pm7.3$ \,rad\,m$^{-2}$,
and a polarisation degree of $7.06 \pm 0.23 \%$.

For the north component we do see a hint of a polarisation emission at $+3.5\pm0.1$\,rad\,m$^{-2}$ which could correspond to the RM found in the jet component in \citet{2009ApJ...702.1230T}.
However, as the source is only partially resolved in the NVSS, it is very difficult to be confident about which RMs correspond to which positions of the source.

Taking the polarised intensity at the Faraday depth $+20.5\pm0.1$\,rad\,m$^{-2}$ we see that the depolarisation factor for the lobe to be 0.196.

While the single lobe is somewhat depolarised at 151\,MHz compared to the 1.4\,GHz value of \citet{2009ApJ...702.1230T}, the core is completely depolarised.
This suggests that the Laing-Garrington effect occurs \citep{1988Natur.331..147G-1}
where the stronger jet is closer to us, is seen through a smaller pathlength of magneto-ionic material, and thus
shows less depolarisation.

\subsubsection{J132626+473741}

\begin{figure*}
\centering
\includegraphics[width=16cm]{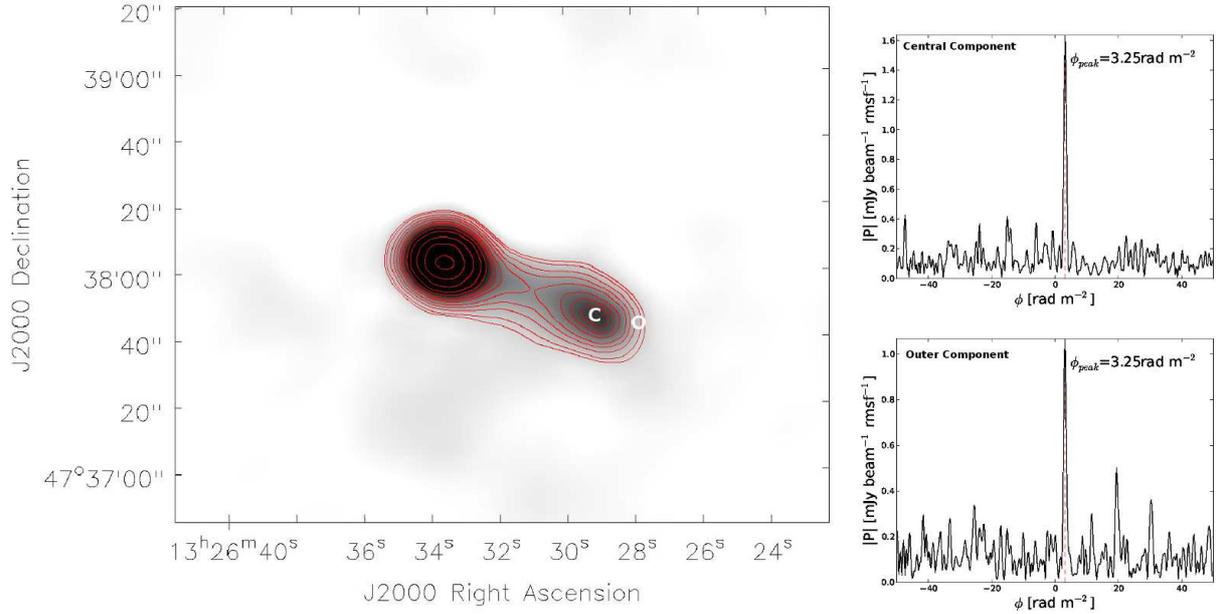}
\caption{The radio lobe J132626+473741 in total intensity (left) with the two marked corresponding Faraday
spectra (right).}
\label{faradayspectra2}
\end{figure*}

\begin{figure*}
\centering
\includegraphics[width=18cm]{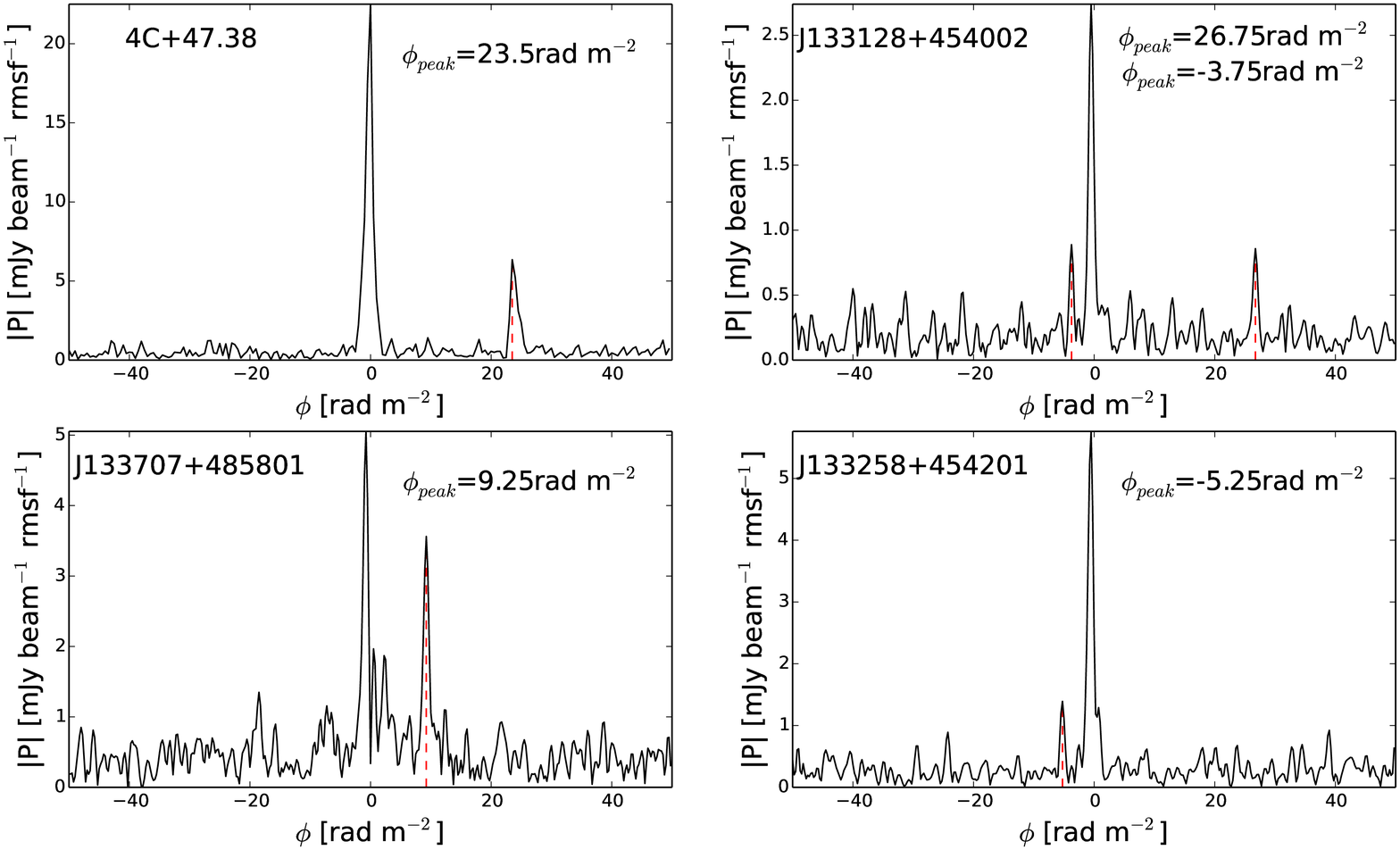}
\caption{The Faraday Spectra of the four unresolved polarised sources detected in the M51 field.}
\label{all4faradayspec}
\end{figure*}

This source is also a single lobe radio galaxy with a strong polarisation detection in the lobe at
$\phi = +3.2\pm0.1\,$rad\,m$^{-2}$. Unfortunately, this source was not detected by \citet{2009ApJ...702.1230T} and
therefore no comparison can be made with higher frequencies.
At the peak of the lobe, the polarisation degree is $2.9\pm0.2\%$, very similar to the previous source J133920+464115,
and has one clear component in the Faraday spectrum.
Closer to the edge of the radio lobe, marked "O" in Fig.~\ref{faradayspectra2}, we detect a possible secondary component
at $\phi = +19.5\,$rad\,m$^{-2}$ with a polarised flux density of 390\,$\mu$Jy/beam and possibly a third one at
$\phi = +30.5\,$rad\,m$^{-2}$ with a polarised flux density of 280\,$\mu$Jy/beam.
The Faraday spectra and total intensity image for this source are shown in Fig.~\ref{faradayspectra2}.

\section{Discussion \label{section9}}

The radial profiles of M51 at 151\,MHz and 1.4\,GHz, which consist of mostly nonthermal synchrotron emission, are fitted by a
larger exponential profile at radii $r \le 10$\,kpc than at $ \ge 10$\,kpc.
As mentioned by \citet{2007A&A...472..785T}, synchrotron emission mimics the distribution of the star
forming regions in the disk. As we showed in this study, this still applies for low frequencies. The sharp break
in star formation rate at about 7\,kpc radius is also seen in synchrotron emission, but is shifted and
flattened at 151\,MHz due to the propagation of low-energy CREs.
As a result, the expectation that galaxies would be extremely large at low frequencies \citep{2013AN....334..548B} does not seem to hold.

Our observations demonstrate that a significant magnetic field is present up to 16\,kpc in the outer disk due
to the fact that we observe synchrotron emission.
The magnetic field must be at least $3.25\,\mu$G in order for synchrotron emission to be visible despite inverse Compton losses.
However, it is very difficult to determine accurately the magnetic field strength in these regions because of the
uncertainty of the CR proton to electron number density in areas of low star formation.

Our results give no indication of a flattening of the integrated spectrum of M51 down to 151\,MHz
and therefore disagree with \citet{1991A&A...250..302P} (their Fig.~11).
\citet{1991A&A...250..302P} and \citet{1991A&A...251..442H} argued that a
steepening in the integrated spectrum
of M51 occurs beyond about 1\,GHz, caused by the energy losses of the CREs.
It is likely that this interpretation was affected by the potential
inaccuracies of a few flux density measurements.

According to \citep{1991A&A...250..302P}, there should be a break $E_\mathrm{b}$ of the
CRE energy spectrum at energy $E_\mathrm{b}$
(and a steepening of the synchrotron spectrum beyond the corresponding frequency
according to Eq.~\ref{equationenergycosmicray}):
\begin{equation}
E_\mathrm{b} = \frac{3.3\,10^{15}\,(\frac{\rm{div} (v)}{\mathrm{s}^{-1}})+ 8\,(\frac{n}{1\,\mathrm{cm}^{-3}})}{(\frac{w(r)}{0.7\,\mathrm{eV}\,\mathrm{cm}^{-3}}) + 0.25\,(\frac{B_{\perp}}{3.25\,\mathrm{\mu G}})^{2}} \mathrm{GeV}\, ,
\label{pohlequation}
\end{equation}
\noindent where $\rm{div} (v)$ is the adiabatic expansion term by a galactic wind, $n$ represents the bremsstrahlung
loss term with $n$ being the total gas density, $w(r)$ is the radiation field representing inverse Compton loss,
and $B_{\perp}^{2}$ represents the synchrotron loss term with $B_{\perp}$ being the magnetic field strength
perpendicular to the line of sight. Inverse Compton loss in the galaxy's radiation field is
generally smaller than synchrotron loss in galaxies \citep{2014AJ....147..103H}.

In M51, no spectral break is observed (Fig.~\ref{integratedfluxpic}).
Neglecting inverse Compton loss
and adiabatic expansion and adopting an average field strength of 15\,$\mu$G (Fig.~\ref{Bfieldpic}),
we would need an average total gas density of more than 7\,cm$^{-3}$ to shift $E_\mathrm{b}$ beyond 10\,GeV, so that no
steepening occurs within the observed frequency range (Fig.~\ref{integratedfluxpic}). Such a large gas density is improbable.
Typical gas densities of a few cm$^{-3}$ lead to $E_\mathrm{b}\approx 2-5$\,GeV, hence
a steepening of the synchrotron spectrum beyond 1--6\,GHz is expected.

The model by \citet{1991A&A...250..302P} did not take into account
thermal absorption and thermal emission. While thermal absorption is not important for the integrated spectrum of M51, thermal emission is significant at high frequencies and should lead to a spectral flattening. We propose that a spectral steepening occurs within the observed frequency range, but is hidden by spectral flattening due to the thermal emission.

Note that even a sharp break in the CRE energy spectrum leads to a
smooth steepening of the synchrotron spectrum
due to the broad emission spectrum of each CRE particle. Density,
radiation field and magnetic field strength in galaxies
vary over some range, so that no sharp break exists and the spectrum of
the integrated synchrotron emission is smoothed
out further. Modelling of galaxies is needed to investigate the amount
of variation of synchrotron spectral index.

Indications of thermal absorption only exist in limited regions in the centre of M51. One such region lies
in a complex of {\sc H\,ii} regions, namely, CCM15, 6A, 25B, 37A and 45B \citep{1969A&A.....1..479C}.
This supports the results
of \citet{2013MNRAS.431.3003L} that {\sc H\,ii} regions are the most important locations
of thermal absorption down to 10\,MHz. If this is the case, we expect to detect increased
thermal absorption with increasing pathlength and therefore with increasing inclination of the galaxy.
Observations of edge-on galaxies with LOFAR are important to confirm this, with NGC891 being observed already.

The spiral arms of M51 are still observable at 151\,MHz but signs of cosmic ray diffusion of low-energy cosmic ray electrons are obvious.
For example, the wavelet cross-correlation between 151\,MHz and 70\,$\mu$m is seen to have a significant correlation
at scales greater than 1.45\,kpc compared to 0.72\,kpc for 1.4\,GHz and 70\,$\mu$m.
This is a measure of the propagation length of cosmic ray electrons that decreases with increasing frequency.
We could also detect a smoother gradient in the arm--interarm contrasts at 151\,MHz compared to higher frequencies.
These findings point to the fact that diffusion is the most dominant process of cosmic ray electron propagation
in the star forming disk of M51. In the Milky Way, a diffusion model with the possible inclusion of convection is
the most suitable description of CR transport in the Galaxy at energies below approximately $10^{17}$\,eV
\citep{2007ARNPS..57..285S}. More evidence of this from observations and simulations will be presented in a
future paper (Mulcahy, Fletcher et al., in prep).

Other face-on galaxies with different properties, such as NGC628 with its extended {\sc H\,i} disk \citep{2008AJ....136.2563W}
need also to be examined. Several more face-on galaxies have been been or will be shortly observed with LOFAR and
will shed more light on this matter.
In addition, dedicated LOFAR Low Band Antenna (LBA) observations at 30--70\,MHz will be proposed where thermal
absorption will become more prominent and therefore a flattening of the integrated spectrum is expected to
be found.

\section{Conclusions \label{section10}}

We have presented the first LOFAR High Frequency Antenna (HBA) image of a nearby galaxy, namely M51 at a
central frequency of 151\,MHz.
LOFAR enables us for the first time to observe this galaxy at such a low-frequency with
arcsec resolution and sub-milliJansky sensitivity. This observation enabled us to study aged electrons within the
galaxy and therefore probe regions where the magnetic field is weak, i.e. the extended outer disk.

The main findings are summarised as follows:

\begin{enumerate}
\item We have been able to detect the disk of M51 out to 16\,kpc radius. Assuming equipartition we estimate that the
magnetic field strength is approximately 10\,$\mu$G at a radius of 10\,kpc.

\item The spectrum of integrated radio emission of M51 is described well by a power law with a constant spectral index between 151\,MHz and 22.8\,GHz. The lack of
flattening towards low frequencies indicate that thermal absorption and ionisation losses are not important.
Observations of more galaxies with LOFAR HBA \& LBA and at frequencies beyond 22\,GHz would help to verify this.

\item The lack of steepening of the integrated spectrum of M51 towards high frequencies suggests that flattening
by thermal emission balances steepening by synchrotron losses of CREs. Bremsstrahlung loss dominates below the break
frequency of a few GHz, which is consistent with an average density of the total gas of a few cm$^{-3}$.

\item We see evidence of free-free absorption in the central region and in {\sc H\,ii} regions in the inner arms of
M51 with spectral indices flatter than $-$0.5.

\item The scale length of the synchrotron emission from the outer disk ($r\ge$ 10\,kpc) is approximately 2.6 times
smaller than the scale length in the inner disk ($r\le$ 10\,kpc) at 151\,MHz and 1400\,MHz.
We assign the small scale length in the outer disk to the sharp decrease of the star formation rate and hence of
the sources of cosmic ray electrons. Under such conditions, detecting the extreme outer disk of galaxies
at low radio frequencies and in turn of magnetic fields is difficult.

\item The scale length of the synchrotron emission at 151\,MHz is approximately 1.6 times larger than that at
1400\,MHz in the inner and outer disk, because the CRE propagation length is larger at lower frequencies.

\item While the spiral arm and interarm regions are very visible in both total power and spectral index, we
see significant CRE diffusion into the interarm region at 151\,MHz compared to 1.4\,GHz and conclude that diffusion
is the dominant process of CRE propagation in the star forming disk of M51.

\item The correlations between the images of radio emission at 151\,MHz and 1400\,MHz and the far-infrared emission at
$70\,\mu$m reveal breaks at scales of 1.4 and 0.7\,kpc, respectively. The ratio of these break scales is also
consistent with diffusive CRE propagation.

\item We estimate the average diffusion coefficient of CREs in M51 as 
$D \simeq 3.3 \times 10^{27}$ cm$^{2}$\,s$^{-1}$, smaller than
the values of $D \sim (2-4) \times 10^{28}$\,cm$^{2}$\,s$^{-1}$ used by
\citet{1998ApJ...493..694M} for the Milky Way.

\item M51 was not detected in polarisation at 151\,MHz due to strong Faraday depolarisation.
Detection of diffuse polarisation in star forming regions of galaxies at such frequencies is difficult.
Targets with significant magnetic
fields but little star formation may offer chances to observe diffuse polarisation at low frequencies.

\item We detect six extragalactic sources in polarisation at 151\,MHz with a resolution of 20 arcsec and a rms noise of 100\,$\mu$Jy/beam/rmsf in polarised intensity. In a field the size of approximately
$4.1\degr \times 4.1\degr$, this yields a number density of polarised sources at 151\,MHz of one for every
1.7 square degrees.

\item Sources with a single lobe are far less depolarised than other sources, which
is probably due to the Laing-Garrington effect. A larger sample and detailed modelling are needed to understand
the depolarisation processes of radio galaxies at low frequencies. Such investigations are crucial for
future low-frequency polarisation surveys like SKA--LOW, with the aim of using RM grids of background polarised sources.
\end{enumerate}

Modelling of CRE propagation using the propagation and energy-loss equation is essential to understand the
observational results of this paper. This is already been addressed and will be followed in a later paper
(Mulcahy, Fletcher et al., in prep). In addition, observations of M51 with the LOFAR Low Band Antenna (LBA)
at frequencies down to 30\,MHz would enable us to observe even further out in the galactic outer disk, and the
properties of low-energy CREs, as well to determine the extent of free-free absorption in the star forming
regions of M51, in particular in the central starforming region.
Additional face-on galaxies have been observed in LOFAR Cycle 0 and Cycle 1. This will enable us to compare
galaxies with different properties, to search for diffuse polarisation in the outer galactic disks and to
learn more about the strength and origin of magnetic fields in these regions.

\begin{acknowledgements}
This research was performed in the framework of the DFG Research Unit 1254 ``Magnetisation of Interstellar and Intergalactic Media: The Prospects of Low-Frequency Radio Observations''.
LOFAR, designed and constructed by ASTRON, has facilities in several countries, that are owned by various parties (each with their own funding sources), and that are collectively operated by the
International LOFAR Telescope (ILT) foundation under a joint scientific policy.
We thank the anonymous referee for useful comments and Uli Klein for discussions of galaxy spectra.
\end{acknowledgements}

\bibliographystyle{bibtex/aa}
\bibliography{bibtex/ref}

\begin{landscape}
\begin{table}
\caption{\label{polarisedtable}
Detections of polarised background sources in the M51 field with DP showing the depolarisation factor for the respective sources}
\begin{tabular}{|c|c c|c c c c|c c c|c|}
\hline
Name & RA & Dec & I [mJy] & $\phi$ [rad/m$^2$] & PI [mJy] & p [\%] & RM [rad/m$^2$] & PI [mJy] & p [\%] & DP\\
  & (J2000) & (J2000)  &  & &  &  &   \multicolumn{3}{c|}{(Taylor et al. 2009)} &  \\
\hline
$J132626+473741$& $13^h26^m28^s.9$ & $47^\circ37^\prime41.82^{\prime \prime}$ & $44.2\pm0.85$  & $+3.2\pm0.1$  & $1.6\pm0.1$  & $2.9\pm0.2$  &  &  &  &  \\
$J133128+454002$ & $13^h31^m28^s.6$ & $45^\circ40^\prime4.8^{\prime \prime}$ & $313\pm6.26$ & $-3.8\pm0.1$ & $0.89\pm0.16$ & $0.28\pm0.05$ &  &  &  & \\
$J133258+454201$ & $13^h32^m58^s.8$ & $45^\circ42^\prime2.7^{\prime \prime}$ & $974\pm19$ & $-5.2\pm0.1$ & $1.4\pm0.2$ & $0.14\pm0.02$ &  &  &  & \\
$J133707+485801$ & $13^h37^m07^s.9$ & $48^\circ58^\prime04.26^{\prime \prime}$  & $1500\pm30$ & $+9.2\pm0.1$  & $3.52\pm0.28$  & $0.23\pm0.02$  & $-8.9\pm3.2$  & $16.9\pm0.3$  & $6.0\pm0.1$ &  0.038 \\
$J133920+464115$ & $13^h39^m19^s.4$ & $46^\circ41^\prime19.82^{\prime \prime}$ & $ 219\pm4.4$ & $+20.5\pm0.1$  & $5.28\pm0.15$ & $2.6\pm0.1$  & $5.5\pm7.3$ & $7.06\pm0.23$ & $7.7\pm0.25$ & 0.196 \\
$B31337+469$ (core) &$13^h 39^m 23^s.06$ & $46^\circ40^\prime08.40^{\prime \prime}$ & & & & & $22.5 \pm3.9$ & $12.66\pm0.23$ & $3.66 \pm 0.07 $ & \\
$4C +47.38$ & $13^h41^m45^s.0$ & $46^\circ57^\prime16.8^{\prime \prime}$ & $3600\pm72$ & $+23.5\pm0.1$ & $6.4\pm0.3$ & $0.17\pm0.01$ & $-30.6\pm1.4$ & $37.25\pm0.24$  & $5.8\pm0.1$ & 0.029 \\
\hline
\end{tabular}
\end{table}
\end{landscape}

\end{document}